\documentclass[apj]{emulateapj}

\newcommand{\vect}[1]{{\mathbf{#1}}}
\newcommand{\mydotfill}{\leaders\hbox to 2pt{\hss.\hss}\hfill\phantom{.}}

\newcommand{\lJ}{\lambda_\mathrm{J}}
\newcommand{\kJ}{k_\mathrm{J}}
\newcommand{\kJzero}{k_{\mathrm{J}0}}

\newcommand{\Msol}{\mbox{$M_{\sun}$}}

\newcommand{\cm}{\mbox{cm}}

\newcommand{\km}{\mbox{km}}
\newcommand{\kms}{\mbox{km}\,\mbox{s}^{-1}}

\newcommand{\g}{\mbox{g}}
\newcommand{\G}{\mbox{G}}
\newcommand{\pc}{\mbox{pc}}

\newcommand{\K}{\mbox{K}}
\newcommand{\yr}{\mbox{yr}}

\newcommand{\ted}{t_\mathrm{ed}}
\newcommand{\deltavar}{$\Delta$-variance }
\newcommand{\solratio}{{E_\mathrm{sol}/E_\mathrm{tot}}}

\slugcomment{draft \today}

\shorttitle{Magnetic field amplification by gravity-driven turbulence}
\shortauthors{Federrath, Sur, Schleicher, Banerjee, Klessen}

\begin{document}

\title{A new Jeans resolution criterion for (M)HD simulations of self-gravitating gas: Application to magnetic field amplification by gravity-driven turbulence}

\author{Christoph~Federrath\altaffilmark{1,2}, Sharanya~Sur\altaffilmark{2}, Dominik~R.~G.~Schleicher\altaffilmark{3,4}, Robi~Banerjee\altaffilmark{2}, Ralf~S.~Klessen\altaffilmark{2}}
\email{chfeder@ita.uni-heidelberg.de}

\altaffiltext{1}{Ecole Normale Sup\'{e}rieure de Lyon, CRAL, 69364 Lyon, France}
\altaffiltext{2}{Zentrum f\"ur Astronomie der Universit\"at Heidelberg, \\Institut f\"ur Theoretische Astrophysik, Albert-Ueberle-Str.~2, 69120 Heidelberg, Germany}
\altaffiltext{3}{ESO Garching, Karl-Schwarzschild-Str.~2, 85748 Garching, Germany}
\altaffiltext{4}{Leiden Observatory, Leiden University, P.O.~Box 9513, NL-2300 RA Leiden, The Netherlands}

\begin{abstract}
Cosmic structure formation is characterized by the complex interplay between gravity, turbulence, and magnetic fields. The processes by which gravitational energy is converted into turbulent and magnetic energies, however, remain poorly understood. 
Here, we show with high-resolution, adaptive-mesh simulations that MHD turbulence is efficiently driven by extracting energy from the gravitational potential during the collapse of a dense gas cloud. Compressible motions generated during the contraction are converted into solenoidal, turbulent motions, leading to a natural energy ratio of $\solratio\approx2/3$. We find that the energy injection scale of gravity-driven turbulence is close to the local Jeans scale. If small seeds of the magnetic field are present, they are amplified exponentially fast via the small-scale dynamo process. The magnetic field grows most efficiently on the smallest scales, for which the stretching, twisting, and folding of field lines, and the turbulent vortices are sufficiently resolved. We find that this scale corresponds to about 30 grid cells in the simulations. We thus suggest a new minimum resolution criterion of 30 cells per Jeans length in (magneto)hydrodynamical simulations of self-gravitating gas, in order to resolve turbulence on the Jeans scale, and to capture minimum dynamo amplification of the magnetic field. Due to numerical diffusion, however, any existing simulation today can at best provide lower limits on the physical growth rates. We conclude that a small, initial magnetic field can grow to dynamically important strength on time scales significantly shorter than the free-fall time of the cloud.

\end{abstract}


\section{Introduction}
Astrophysical fluids on virtually all scales are characterized by highly complex turbulent motions. Besides being turbulent, we know from observations that the gas is magnetized between galaxies and inside galaxies, as well as in individual star-forming molecular clouds, down to the proto-stellar accretion disks, which naturally accompany the birth of stars and planetary systems. However, not just the present-day universe is characterized by turbulence and magnetic fields. With the advent of large-scale computer simulations, it became clear that primordial gas becomes highly turbulent during the formation of the first galaxies and stars as well \citep[e.g.,][]{AbelBryanNorman2002,OSheaNorman2007,WiseAbel2007,ClarkGloverKlessen2008,GreifEtAl2008}, which has far-reaching consequences for all cosmic structure formation. This turbulence provides an extremely efficient way to amplify primordial seeds of the magnetic field \citep[e.g.,][]{Biermann1950,KulsrudEtAl1997,LazarEtAl2009,XuEtAl2009,MiniatiBell2011} by the turbulent dynamo process \citep[see the review by][]{BrandenburgSubramanian2005} already during the earliest phases of structure formation in the universe \citep{RyuEtAl2008}. Here the amplification of the magnetic field arises from sequences of a ``stretch, twist, and fold'' nature, until the magnetic field lines are so tightly packed that the magnetic energy density becomes comparable to the kinetic energy density. Strong magnetic fields, close to equipartition may have been reached in normal disk galaxies already at redshifts of $z\lesssim3$ \citep[][and references therein]{BernetEtAl2008,ArshakianEtAl2009}, emphasizing the significance of the dynamo action even in the early stages of galaxy evolution and star formation. Thus, magnetic fields may not only have a significant impact on the gas dynamics and fragmentation in present-day star formation \citep{HennebelleTeyssier2008}, but potentially also in primordial star formation.

Despite its ubiquity and importance for star formation \citep[][]{MacLowKlessen2004,ElmegreenScalo2004,McKeeOstriker2007}, only very little is known about the origin of astrophysical turbulence and magnetic fields. Recently, \citet{KlessenHennebelle2010}, \citet{ElmegreenBurkert2010}, and \citet{VazquezSemadeni2010} suggested that a universal source for driving astrophysical turbulence everywhere from cosmic scales and galaxies, over galactic clouds, down to individual proto-stellar disks is the accretion from their environment, i.e., by gravity-driven flows, on all scales \citep[see also,][]{FieldBlackmanKeto2008}. Similar ideas existed earlier. In particular, \citet{Hoyle1953} showed that during the contraction of a slightly unstable, nearly isothermal gas cloud, internal random motions can be excited. \citet{OdgersStewart1958} suggested that even fully irrotational motions can give rise to turbulence due to the Reynolds stresses. Hoyle's idea of gravity-driven turbulence was later refined by \citet{ScaloPumphrey1982}, which they called `turbulent virialization'. \citet{Fleck1983} suggested that the injection of turbulence by gravitational contraction is important in the interstellar medium, and \citet{BegelmanShlosman2009} conclude that angular-momentum transport during the turbulent collapse of a gaseous system may suppress fragmentation.

Here we show with numerical simulations that turbulence and magnetic field growth are indeed efficiently driven by the gravitational energy released during the collapse of a dense gas cloud. The connection between gravity-driven turbulence and magnetic field amplification has also been suggested recently in a model by \citet{SchleicherEtAl2010} and confirmed numerically in \citet{SurEtAl2010}. In this process, potential energy is converted into turbulent motions, which in turn amplify the magnetic energy via the turbulent dynamo. Thus, the driving of turbulence and magnetic field growth by gravitational infall may be the consequence of a self-regulating instability. At the bottom of this cascade, close to the sonic scale \citep{FederrathDuvalKlessenSchmidtMacLow2010}, gas is expected to become subsonic as a consequence of a steep rise in the temperature when the gas becomes optically thick. During this process, compressible modes will be converted into solenoidal turbulent motions, until a natural energy ratio of $\solratio\approx2/3$ is reached, \citep{ElmegreenScalo2004,FederrathKlessenSchmidt2008}.

In this paper, we first discuss the physics of magnetic field amplification by gravity-driven turbulence, and subsequently derive a new resolution criterion required to resolve these processes. We aim to address the following key questions: On what scales does the magnetic field grow during the collapse of a dense, magnetized gas cloud? What is the effective kinetic energy injection scale of gravity-driven turbulence? How are the compressible motions in a contracting system converted into turbulent random motions, and what is the asymptotic fraction of solenoidal motions generated during the contraction? After presenting our methods in section~\ref{sec:methods}, we address these physical questions in section~\ref{sec:gravitydriventurb} with the use of magnetohydrodynamical (MHD) simulations. We find that the magnetic field is most efficiently amplified on the smallest resolvable scales in the simulations, and grows exponentially fast due to the small-scale dynamo. The effective energy injection scale of gravity-driven turbulence is close to the local Jeans scale during the contraction. Finally, we show in section~\ref{sec:rot_ratio} that about 2/3 of the total kinetic energy released during the collapse is converted into solenoidal, turbulent motions, efficiently driving magnetic field amplification.

In the second part of the paper (section~\ref{sec:newjeansresol}), we discuss the numerical resolution requirements for modeling turbulent, self-gravitating systems and for minimum dynamo action to set in. To study the resolution dependence of our results, we use a sequence of simulations in which we resolve the Jeans length with 8, 16, 32, 64, and 128 grid cells \citep[see,][paper I]{SurEtAl2010}. With a Fourier analysis of the magnetic energy, we confirm our earlier findings in paper I, showing that 30 grid cells per Jeans length is the threshold resolution for minimum dynamo amplification of the magnetic field to occur. In addition, we show that a Jeans resolution of about 30 grid cells is required to obtain converged values of the turbulent energy, in particular of the solenoidal (rotational) component, which drives dynamo amplification.

In contrast, we find that the turbulent energy density, i.e., the turbulent pressure on the Jeans scale is underestimated, if the Jeans length is resolved with 16 grid cells or less during the collapse. Apart from a few exceptions \citep[e.g.,][]{AbelBryanNorman2002}, the Jeans length is resolved with less than 16 grid cells in almost all numerical simulations today. We speculate that this is because in the study by \citet{TrueloveEtAl1997}, it was found that only four grid cells per Jeans length are enough to avoid artificial fragmentation. An equivalent resolution criterion for the Jeans mass in smoothed particle hydrodynamics (SPH) simulations was found by \citet{BateBurkert1997}. Also, the computational expenses increase strongly, if one aims to resolve the Jeans length with more than a few cells (or few particles in SPH). Thus, most existing hydrodynamical and MHD simulations of self-gravitating media have typically used about ten grid cells per Jeans length or less. Some modification of this criterion was recently also suggested by \citet{GawryszczakEtAl2010} to better resolve self-gravitating disks.

Here we find that the turbulence and magnetic field structure are under-resolved, if the Jeans length is resolved with 16 cells or less. Moreover, turbulent dynamo amplification of the magnetic field is completely absent in this case. To avoid this problem, we suggest a new resolution criterion for simulations of self-gravitating gaseous media: in order to resolve turbulence on the Jeans scale and to account for the turbulent pressure on that scale, as well as to capture minimum dynamo action in MHD simulations, 30 grid cells per Jeans length are required, which is discussed in detail in section~\ref{sec:newjeansresol}.

\section{Method} \label{sec:methods}

Gravitational collapse, turbulence and magnetic field evolution are --except for some idealized cases-- difficult to study analytically, because the system is highly non-linear and naturally three-dimensional. Thus, we study the processes leading to turbulence and magnetic field amplification through gravitational collapse in high-resolution MHD simulations.

\subsection{Initial conditions and setup of our MHD simulations of a collapsing dense core}
We present results of a numerical experiment with a collapsing, magnetized, turbulent gas core \citep[see,][paper I]{SurEtAl2010}. We focus on the gravitational collapse and magnetic field amplification of a dense gas cloud, using a simplified setup, where we assume an almost isothermal equation of state (effective $\Gamma=d\log T/d\log\rho+1=1.1$, with the temperature $T$ and density $\rho$) and neglect non-ideal MHD effects (discussed below). The numerical simulations presented here were performed with the publicly available adaptive-mesh refinement (AMR) code, FLASH2.5 \citep{FryxellEtAl2000}. We solve the equations of ideal MHD, including self-gravity with a refinement criterion guaranteeing that the Jeans length, 
\begin{equation}
\label{eq:jlength}
\lJ = \left(\frac{\pi\,c_{\mathrm s}^{2}}{G\,\rho}\right)^{1/2}\,,
\end{equation}
with the sound speed $c_{\mathrm s}$, gravitational constant $G$, and density $\rho$, is always resolved with a user-defined number of cells. We use the new HLL3R scheme for ideal MHD \citep{WaaganFederrathKlingenberg2011}, which employs a 3-wave approximate MHD Riemann solver \citep{BouchutKlingenbergWaagan2007,Waagan2009,BouchutKlingenbergWaagan2010}. This MHD scheme is an extension of the hydrodynamical version \citep{KlingenbergSchmidtWaagan2007}, developed for FLASH that preserves physical states (e.g., positivity of mass density and pressure) by construction, and is highly efficient and accurate in modeling astrophysical MHD problems involving turbulence and shocks \citep{WaaganFederrathKlingenberg2011}.

The ionization degree is expected to be sufficiently high in primordial clouds to ensure a strong coupling between ions and neutrals to maintain flux-freezing \citep{MakiSusa2004}. However, non-ideal MHD effects may eventually become important at very high densities, as suggested by simulations of contemporary star formation \citep[e.g.,][]{HennebelleTeyssier2008,DuffinPudritz2009}. These effects are not included in the present calculations, but should be subject of future studies.

The initial conditions for our numerical simulations were motivated from large-scale cosmological models \citep[e.g.,][]{AbelBryanNorman2002,YoshidaOmukaiHernquist2008}. However, for simplicity we use a spherically symmetric initial density distribution, following a super-critical Bonnor-Ebert profile \citep{Ebert1955,Bonnor1956}  with a core density of $\rho_{\mathrm BE}=4.68\times10^{-20}\,\g\,\cm^{-3}$ at a temperature of $T=300\,\K$. We did not include any initial rotation of the core. The Bonnor-Ebert sphere has a radius of $1.5\,\pc$ and a total mass of $M=1500\,\Msol$. The three-dimensional Cartesian computational domain has a volume of $(3.9\,\pc)^3$. For a similar approach see \citet{ClarkGloverKlessen2008,ClarkGloverKlessenBromm2011}

Although the existence of magnetic fields on a wide range of astrophysical scales is clear, their origin is still poorly understood. Some mechanisms include the \citet{Biermann1950} battery, the Weibel instability \citep[e.g.,][]{LazarEtAl2009}, the ejection of magnetic fields from active galactic nuclei \citep[e.g.,][and references therein]{XuEtAl2009}, production by cosmic rays \citep[][]{MiniatiBell2011}, and amplification by galaxy cluster mergers. Recent lower limits of magnetic fields in the intergalactic medium suggest $B_\mathrm{rms}\gtrsim10^{-15}\,\mathrm{G}$ \citep{TavecchioEtAl2010,DolagEtAl2011}. To keep our setup as simple and general as possible, we add a small initial seed of the magnetic field and study its self-consistent amplification during the collapse of a gravitationally unstable gas cloud in detail. In order to initialize seeds of turbulence and magnetic field, we impose a random initial velocity field with transonic velocity dispersion of amplitude $1.1\,\km\,\mathrm{s}^{-1}$ (equal to the initial sound speed) and a weak random magnetic field with $B_\mathrm{rms}=1\,\mathrm{nG}$, without a mean field component. This initial field strength corresponds to a plasma beta (ratio of thermal to magnetic pressure) of $\beta\approx10^{10}$. Both, the turbulent velocity and magnetic field were set up with the same power-law dependence in wavenumber space, $P(k)\propto k^{-2}$, peaking on scales of about $0.8\,\pc$, which roughly corresponds to the initial Jeans length of the core. The choice of the initial power spectrum is motivated by the finding that turbulence develops a $k^{-5/3}$ and $k^{-2}$ velocity spectrum for incompressible \citet{Kolmogorov1941c} and for shock-dominated \citet{Burgers1948} turbulence, respectively. Astrophysical systems typically exhibit spectra of the turbulence in between these two extremes. Since both, Kolmogorov and Burgers turbulent spectra are dominated by large-scale modes, i.e., small wavenumbers $k$, our results are not expected to depend strongly on the particular choice of the initial spectrum. However, varying parameters of the initial conditions will be subject of a follow-up study. After an initial transient phase, the turbulence develops its own self-consistent spectrum, driven by the gravitational collapse.

We consider an isolated system for solving the Poisson equation, and use Neumann boundary conditions (zero-gradient) for the MHD. Since the initial conditions for the magnetic field were generated in Fourier space to obtain a divergence-free field, the initial magnetic field is naturally periodic. This introduces an inconsistency at the six boundaries of the simulation domain and may cause some non-vanishing divergence there. However, even the fastest MHD waves (the fast magneto-sonic waves) have a box-crossing time of about $3\times10^6\,\yr$, an order of magnitude longer than the initial free-fall time scale of the system. Thus, any potential perturbations produced at the boundaries can be ignored in the dense core.

Consistent with previous works of \citet{OmukaiEtAl2005} and \citet{GloverSavin2009}, we adopt an effective equation of state with $\Gamma=1.1$ for number densities in the range $n=10^5$--$10^{10}\,\cm^{-3}$. However, this kind of setup is sufficiently general to allow us to investigate basic properties of turbulence and magnetic field amplification in the context of collapsing primordial as well as present-day gas clouds.

\subsection{Five simulations with increasing Jeans resolution} \label{sec:resolutionstudy}
We note that the efficiency of the turbulent dynamo depends on the kinematic Reynolds number, $\mathrm{Re}=\ell_\mathrm{int}\sigma_v/\nu$, with the driving scale $\ell_\mathrm{int}$, turbulent velocity dispersion $\sigma_v$ on that length scale, and physical viscosity $\nu$. Also note that previous studies found a critical value of the magnetic Reynolds number, $\mathrm{Rm}=\ell_\mathrm{int}\sigma_v/\eta$, below which the small-scale dynamo is not excited \citep{BrandenburgSubramanian2005}. In numerical simulations, however, the viscosity, $\nu$ and the magnetic diffusivity, $\eta$ are typically determined by the numerical cutoff scale, as a result of the discretization of the fluid variables on a computational grid with finite resolution. The numerical values of $\mathrm{Re}$ and $\mathrm{Rm}$ are thus related to the grid resolution and to how well the turbulent motions are resolved \citep[e.g.,][]{HaugenBrandenburgDobler2004,BalsaraEtAl2004,SchekochihinEtAl2004}. Even if physical viscosity/diffusivity is added as an extra term in the equations, it must be guaranteed that it is at least as high as the numerical viscosity/diffusivity. Typical Reynolds numbers reached in numerical simulations of turbulence are of the order of a few hundred \citep[depending on what scales are estimated to be affected by numerical viscosity, which varies among different codes, see][]{KitsionasEtAl2009}, which is usually orders of magnitude below the physical Reynolds numbers in the systems that one aims to study, but just large enough to excite the turbulent small-scale dynamo. A strong resolution dependence of turbulent and magnetic properties is thus expected, with higher resolution resulting in smaller numerical viscosity and diffusivity, higher Reynolds numbers, and thus yielding faster magnetic field amplification. To demonstrate this effect, we perform five numerical simulations, in which we resolve the Jeans length, $\lJ$ with 8, 16, 32, 64, and 128 cells. In our highest resolution simulation with 128 cells per Jeans length, we use AMR with 19 levels of refinement by a factor of two. This results in an effective resolution of $524\,288^3$ grid cells by the time when we have to stop the simulation, because it become computationally prohibitive.

\subsection{Analysis in the collapsing frame of reference} \label{sec:methods_analysis}
To understand the behavior of the system quantitatively, we need to follow its dynamical contraction in an appropriate frame of reference. First, we note that the physical time scale becomes progressively shorter during the collapse. We therefore define a dimensionless time coordinate $\tau$ (see paper I),
\begin{equation} \label{eq:tau}
\tau = \int \frac{1}{t_\mathrm{ff}(t)}\,dt\,,
\end{equation}
which is normalized in terms of the local free-fall time, 
\begin{equation}
t_\mathrm{ff}(t)=\left(\frac{3\pi}{32\,G\langle\rho(t)\rangle}\right)^{1/2},
\end{equation}
where $\langle\rho(t)\rangle$ is the mean density in the central Jeans volume, $V_\mathrm{J}=4\pi(\lJ/2)^3/3$. If not otherwise stated, we obtain all dynamical quantities of interest within this contracting Jeans volume, which is centered on the position of the maximum density. This approach enables us to study the turbulence and magnetic field amplification in the collapsing frame of reference. However, we also analyze radial profiles and magnetic field spectra in the fixed frame of reference to study the global evolution of the system.

\subsection{Fourier analysis} \label{sec:methods_fourier}
To study the scale-dependence of the magnetic field, we use Fourier analysis and compute magnetic energy spectra, $P(B)$, defined as
\begin{equation} \label{eq:mag_spect}
P(B)\,dk = \frac{1}{2}\int\widehat{\vect{B}}\cdot\widehat{\vect{B}}^{*}\,4\pi k^2\,dk\,,
\end{equation}
where $\widehat{\vect{B}}$ denotes the Fourier transform of the magnetic field. The spectral energy density is averaged over spherical shells in Fourier space, such that the directional information of the wavevector is integrated out, and the pure scale-dependence of the field can be studied as a function of the norm of the wavevector, $k=|\vect{k}|$. Note that integrating $P(B)$ over all wavenumbers yields the total magnetic energy. Thus, $P(B)\,dk$ is the magnetic energy on scales between $k$ and $k+dk$.

For computing the spectra inside the Jeans volume of the core, we extract the AMR data inside a cube with length equal to the local Jeans length of the core and interpolate all relevant quantities for the analysis (density, velocity, and magnetic field components) to a uniform grid with resolution equivalent to the local Jeans resolution.

A problem in applying Fourier analysis to non-periodic datasets is that the spectrum can be distorted by the fact that the data is discontinuous at the boundary of the extracted cube. Methods to avoid this include applying a spherical window function, which smoothly approaches zero on all boundaries of the cube, zero-padding around the boundaries, or mirroring the dataset on all sides. We compared spectra obtained by simply ignoring the non-periodicity of the data with spectra obtained by windowing with a spherical Hann window, by zero-padding \citep[e.g.,][]{BruntEtAl2003}, or by mirroring the dataset \citep[e.g.,][]{OssenkopfKripsStutzki2008a}. We find that the spectra are very similar in all cases. They are mostly affected at small scales (high-$k$) when different methods are applied. Since this high-wavenumber range is also strongly affected by numerical resolution, we simply ignore the non-periodicity of the data and compute spectra as if the extracted data were periodic. We also tried a wavelet transformation, the \deltavar technique \citep{StutzkiEtAl1998,OssenkopfKripsStutzki2008a,OssenkopfKripsStutzki2008b} to compute the scale dependence of the magnetic field, and found consistent results. However, since the direct spectral method allows a decomposition of the turbulent velocity field into solenoidal and compressible modes, and thus enables us to study these components separately, we prefer the Fourier analysis for the present purpose.

\section{Magnetic field amplification by gravity-driven turbulence} \label{sec:gravitydriventurb}

\subsection{Time evolution}
The time evolution of the density, magnetic field strength and Mach number of the magnetized, turbulent collapse of our Bonnor-Ebert sphere is discussed in detail in paper I. In particular, we showed that the root-mean-squared (rms) magnetic field increases rapidly from a small seed field of $10^{-9}\,\mathrm{G}$ to the Milli-Gauss level, i.e., over six orders of magnitude in our highest resolution run with 128 cells per Jeans length. We showed that two orders of magnitude of this growth are due to twisting and folding of magnetic field lines, i.e., due to the action of the turbulent small-scale dynamo \citep{BrandenburgSubramanian2005}. This is, however, a lower limit to the physical amplification by the dynamo, as the dynamo growth rate depends strongly on the Reynolds number, and thus on numerical resolution \citep[see, e.g.,][]{HaugenBrandenburgDobler2004}. If the Jeans length is resolved with 16 cells or less \citep[which is quite common in typical present-day grid simulations involving gravity, and correspondingly in simulations with the smoothed particle hydrodynamics technique, see, e.g.,][]{FederrathBanerjeeClarkKlessen2010}, no dynamo amplification is seen at all. Even in our highest resolution run, where we resolve the Jeans length with 128 cells, most of the amplification is due to compression of magnetic field lines, i.e., flux-freezing, and not due to the dynamo. The rather small turbulent Reynolds numbers achievable in modern simulations (of the order of a few hundred), do not allow for a more efficient dynamo amplification in the simulations. However, since the Reynolds numbers in real astrophysical systems are typically much higher (e.g., of the order of $10^7$ in the interstellar medium), dynamo amplification will eventually dominate over compressional amplification.

\subsection{Radial profiles} \label{sec:radprofiles}

\begin{figure}[t]
\centerline{\includegraphics[width=1.0\linewidth]{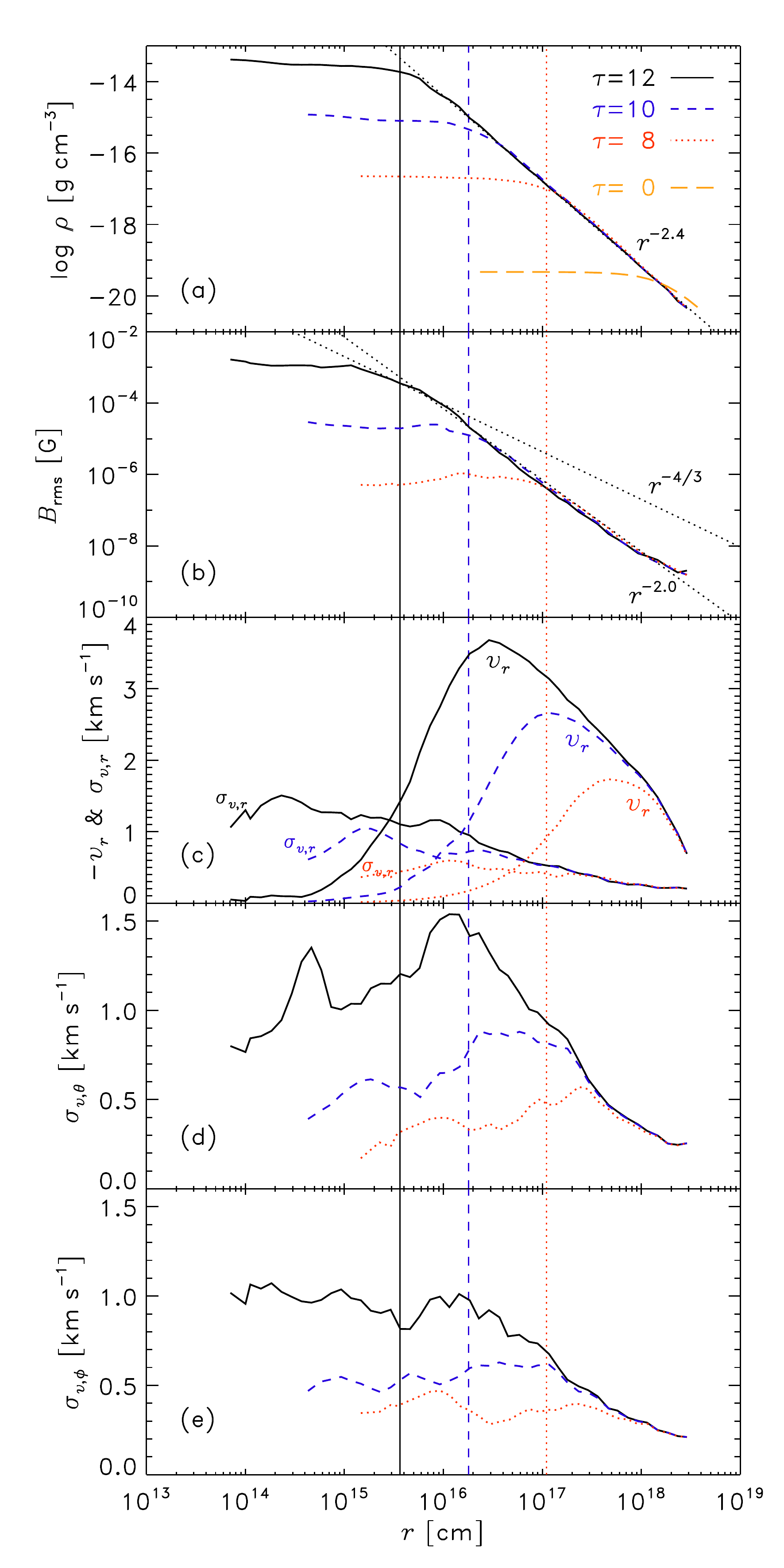}}
\caption{a) Radial density profile of the collapsing, magnetized, turbulent Bonnor-Ebert sphere at times $\tau=8$, 10, and 12. b) Same as a) but for the rms magnetic field. c) Same as a) but for the infall velocity, $-v_r$ and the radial velocity dispersion, $\sigma_{v,r}$. d) and e) Same as a) but for the polar and azimuthal velocity dispersions, $\sigma_{v,\theta}$ and $\sigma_{v,\phi}$. The vertical lines extending through all panels indicate the Jeans radius at $\tau=8$, 10, and 12, respectively. See paper I for the corresponding radial profiles of the rms velocity and Mach number.}
\label{fig:radprofiles}
\end{figure}

Figure~\ref{fig:radprofiles} shows the radial profiles of gas density, rms magnetic field, radial infall velocity, $v_r$ and dispersion, $\sigma_{v,r}$, polar and azimuthal velocity dispersions, $\sigma_{v,\theta}$ and $\sigma_{v,\phi}$ at $\tau=8$, 10, and 12 in the collapse regime for our run with 128 cells per Jeans length (note the definition of the dimensionless time coordinate $\tau$ in eq.~\ref{eq:tau}). The density and radial velocity profiles follow the typical collapse profiles of an unstable gas cloud \citep[see, e.g.,][]{Larson1969,Penston1969}. The initial density profile exhibits a short power law with exponent -2.2 for large radii \citep{Ebert1955,Bonnor1956}, while the profiles at later times are very well fit with a power law, $\rho\propto r^{-2.4}$. This steepening of the profile during the collapse is a result of the slight deviations from an isothermal equation of state (effective $\Gamma=1.1$ instead of 1). The turbulent velocity fluctuations, measured in terms of the dispersions, $\sigma_{v,r}$, $\sigma_{v,\theta}$, and $\sigma_{v,\phi}$ remain subsonic inside the central Jeans volume for $\tau\lesssim12$, because the sound speed increases during the collapse. The rms magnetic field profile exhibits some similarities to the density profile, however, the power-law exponent in the envelope is different. For the 128 cell run, $B_\mathrm{rms}\propto r^{-2.0}$ outside the central Jeans volume, which is significantly steeper than what is expected for purely compressional amplification, i.e., flux-freezing during the collapse ($B_\mathrm{rms}\propto r^{-4/3}$). The steepening is thus due to dynamo amplification. However, since the turbulent dynamo is highly resolution-dependent, the radial power-law exponent is expected to depend on resolution, which we discuss next.
\begin{figure}[t]
\centerline{\includegraphics[width=1.0\linewidth]{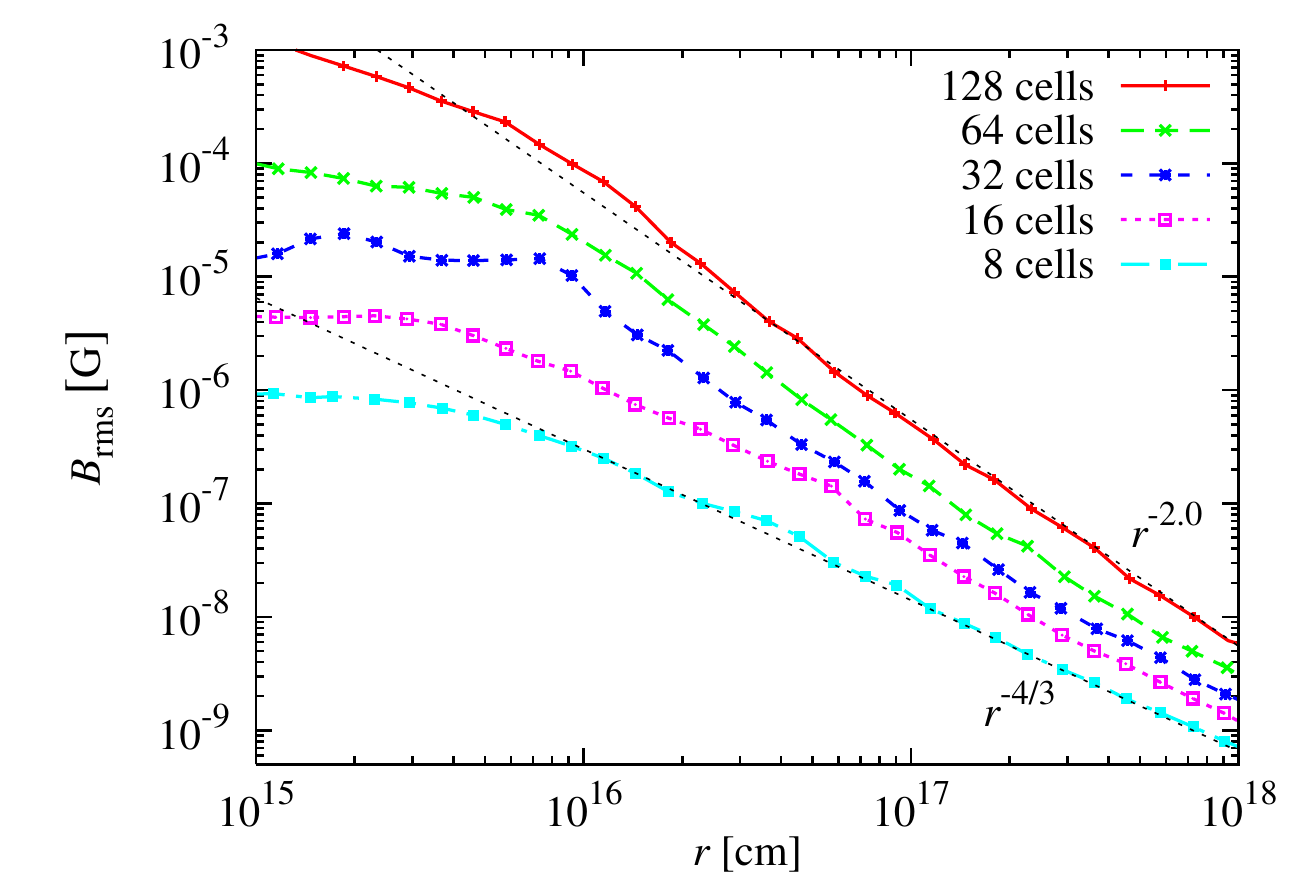}}
\caption{Power-law part of the radial profiles of the rms magnetic field at $\tau=12$ for Jeans resolutions of 8, 16, 32, 64, and 128 cells. Power laws corresponding to pure flux-freezing, $B_\mathrm{rms}\propto r^{-4/3}$, and a steeper, $\propto r^{-2.0}$ profile are drawn for comparison.}
\label{fig:resradprofiles}
\end{figure}

In Figure~\ref{fig:resradprofiles} we show the power-law part of the radial profiles of the rms magnetic field, $B_\mathrm{rms}$ for Jeans resolutions of 8, 16, 32, 64, and 128 cells at $\tau=12$. Both the 8 and 16 cell runs are consistent with the radial exponent expected for pure flux-freezing, $B_\mathrm{rms}\propto r^{-4/3}$. Only for the runs with 32, 64, and 128 cells, we see a steepening of the profiles, which is caused by additional amplification of the field through the turbulent dynamo, again showing that the Jeans length must be resolved with significantly more than 16 cells (e.g., 32 grid cells) for minimum dynamo action to occur. The radial exponent for the 128 cell runs is close to -2.0, however, the profile is expected to steepen further with increasing Jeans resolution. The resolution dependence of our results is discussed in detail in section~\ref{sec:newjeansresol}, where we present evidence for the requirement of a new Jeans resolution criterion in simulations of self-gravitating gas.

Combining the radial dependence of the density, $\rho\propto r^{-2.4}$, and magnetic field, $B_\mathrm{rms}\propto r^{-2.0}$ in our 128 cell run, we find $B_\mathrm{rms}\propto\rho^{0.83}$, which is expected to increase further at higher Reynolds numbers (higher resolution). This is clearly steeper than the pure flux-freezing case, $B_\mathrm{rms}\propto\rho^{2/3}$. Moreover, \citet{SchleicherEtAl2009} showed that already a scaling of $B_\mathrm{rms}\propto\rho^{0.6}$ can lead to a significant change in the thermodynamics of primordial gas, which we show here will have an even stronger effect, when turbulent dynamo amplification is taken into account.

The physical exponent of the radial distribution of the magnetic field is set by the physical viscosity and magnetic diffusivity. Since the physical viscosity and diffusivity can be very small, resulting in Reynolds numbers orders of magnitude higher than what we can model in a computer simulation, the radial profile of the magnetic field is expected to be significantly steeper than $B_\mathrm{rms}\propto r^{-2.0}$. This means that the potential influence of the magnetic field can be significant not only inside the Jeans volume, but also on scales larger than the local Jeans length, because of the strong amplification of the field when the core went through the previous dynamo amplification on scales outside the Jeans volume. 

If the Reynolds numbers are sufficiently high inside the Jeans core, we would expect that the magnetic field can increase to about 10\% of equipartition with the turbulent energy on time scales much shorter than the free-fall timescale, boosting the magnetic field to a dynamically significant level even before the core has had time to contract much. In the saturated phase, the radial dependence of the magnetic field is expected to change. The saturation behavior, however, cannot be addressed with the current simulations, because even at the last available time step (after which the simulation becomes computationally too expensive to advance any further), at $\tau=12$ in the 128 cell run, the ratio of magnetic to kinetic energy in the core is $E_\mathrm{mag}/E_\mathrm{kin}\approx2.2\times10^{-5}$, still far away from the expected saturation  level ($E_\mathrm{mag}/E_\mathrm{kin}\approx10\%$). Thus, the saturation behavior needs to be investigated in a follow-up study with initial field strengths closer to equipartition.

\subsection{Turbulence and magnetic field morphology}

\begin{figure*}[t]
\begin{center}
\def\arraystretch{0.5}
\begin{tabular}{cc}
\includegraphics[width=0.4\linewidth]{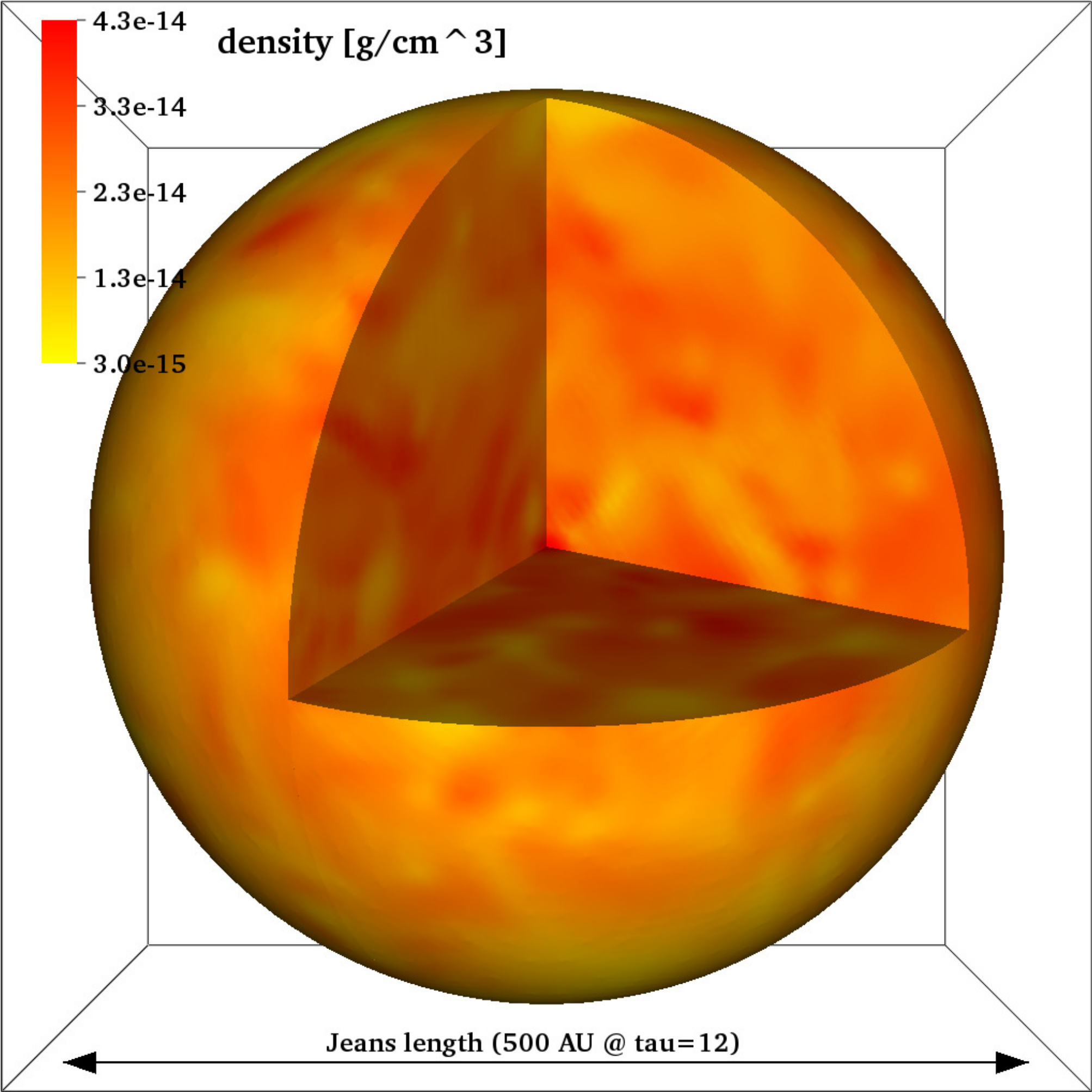} &
\includegraphics[width=0.4\linewidth]{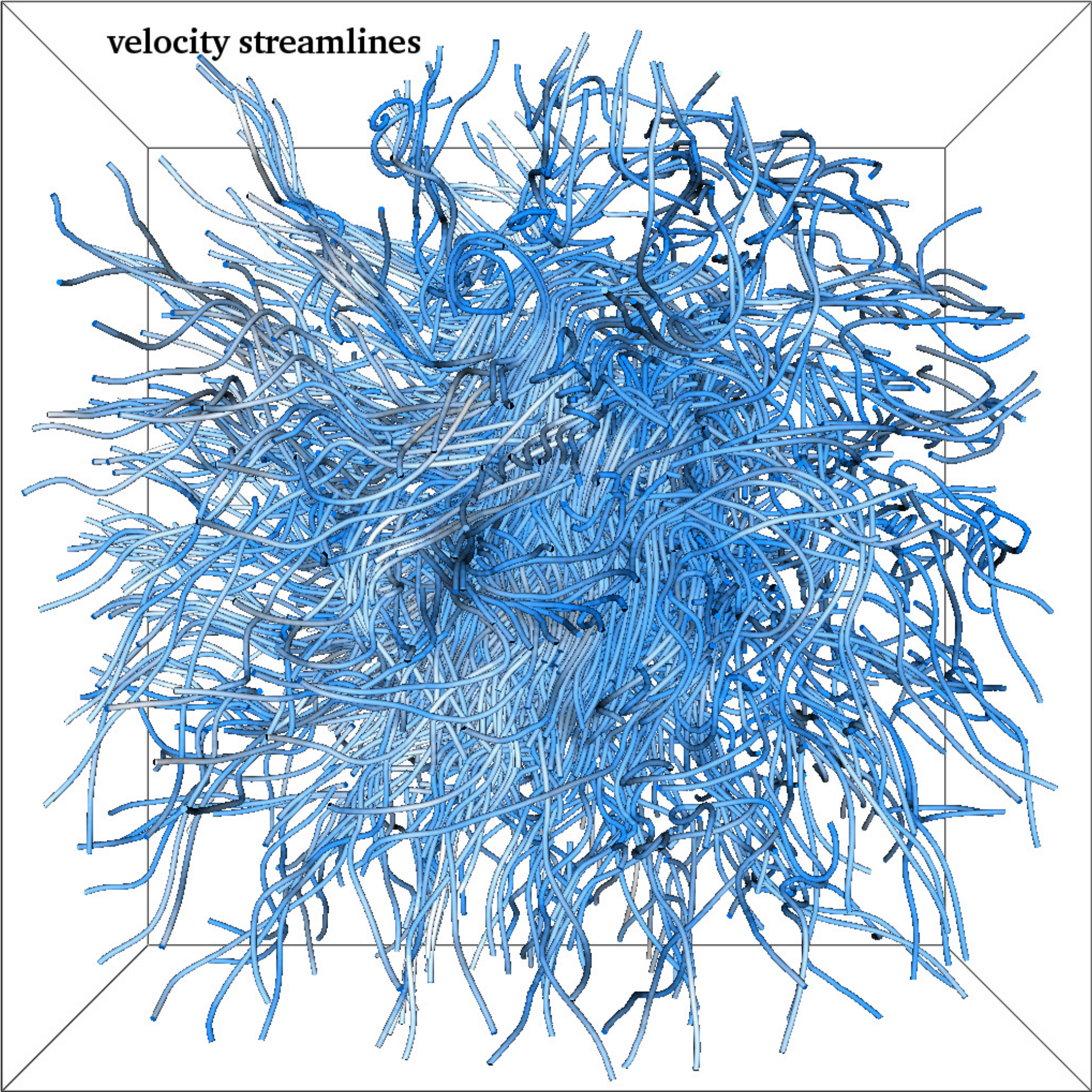} \\
\includegraphics[width=0.4\linewidth]{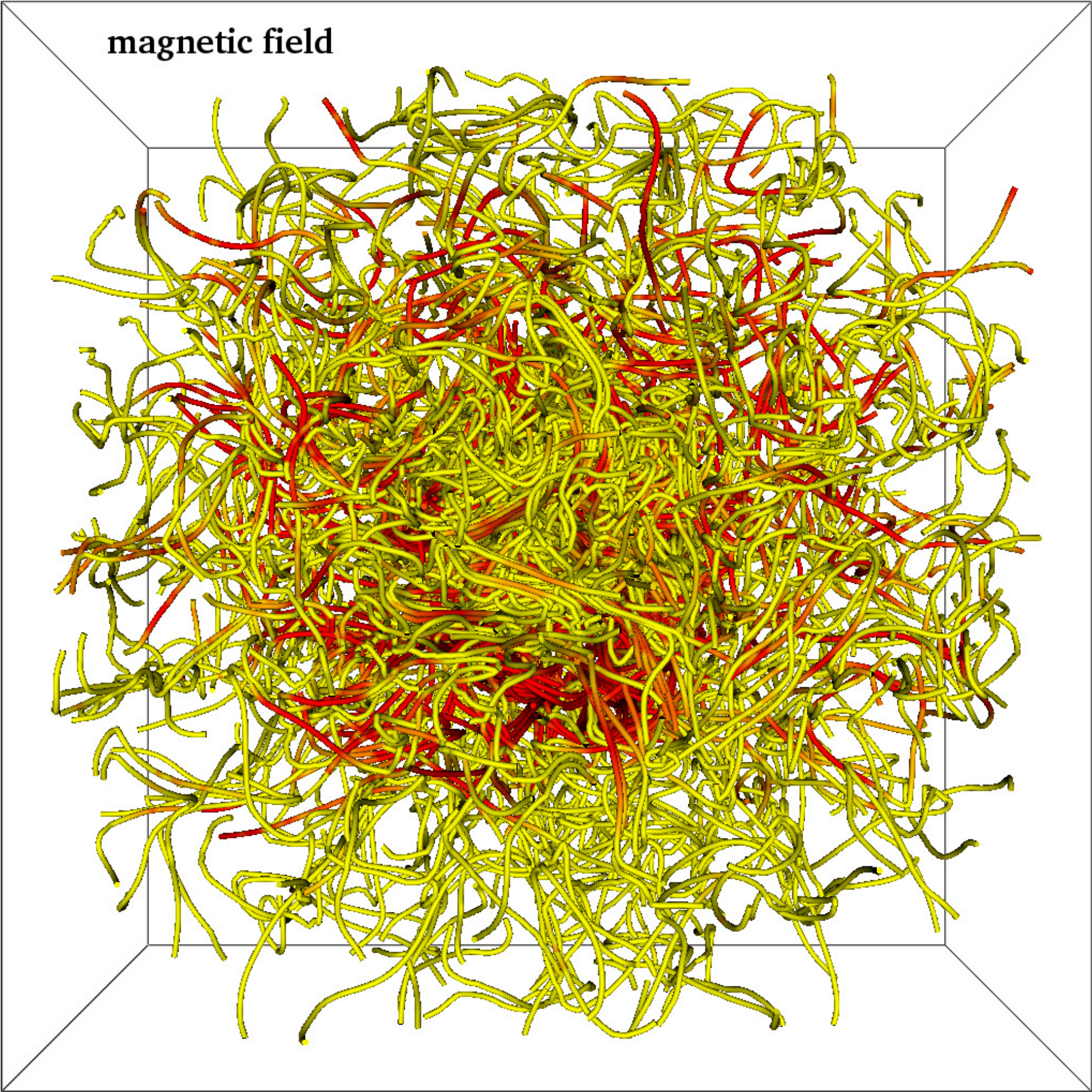} &
\includegraphics[width=0.4\linewidth]{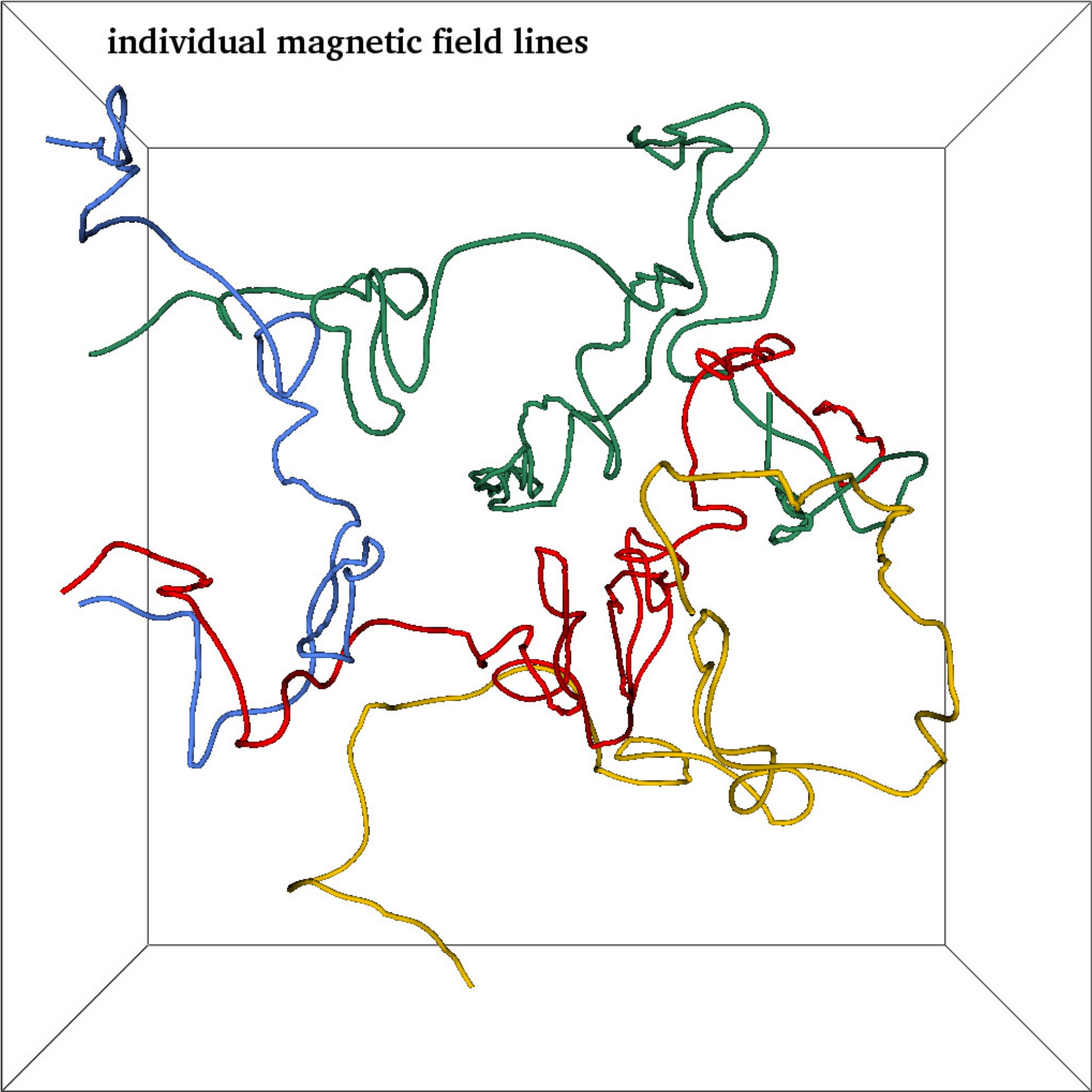} \\
\includegraphics[width=0.4\linewidth]{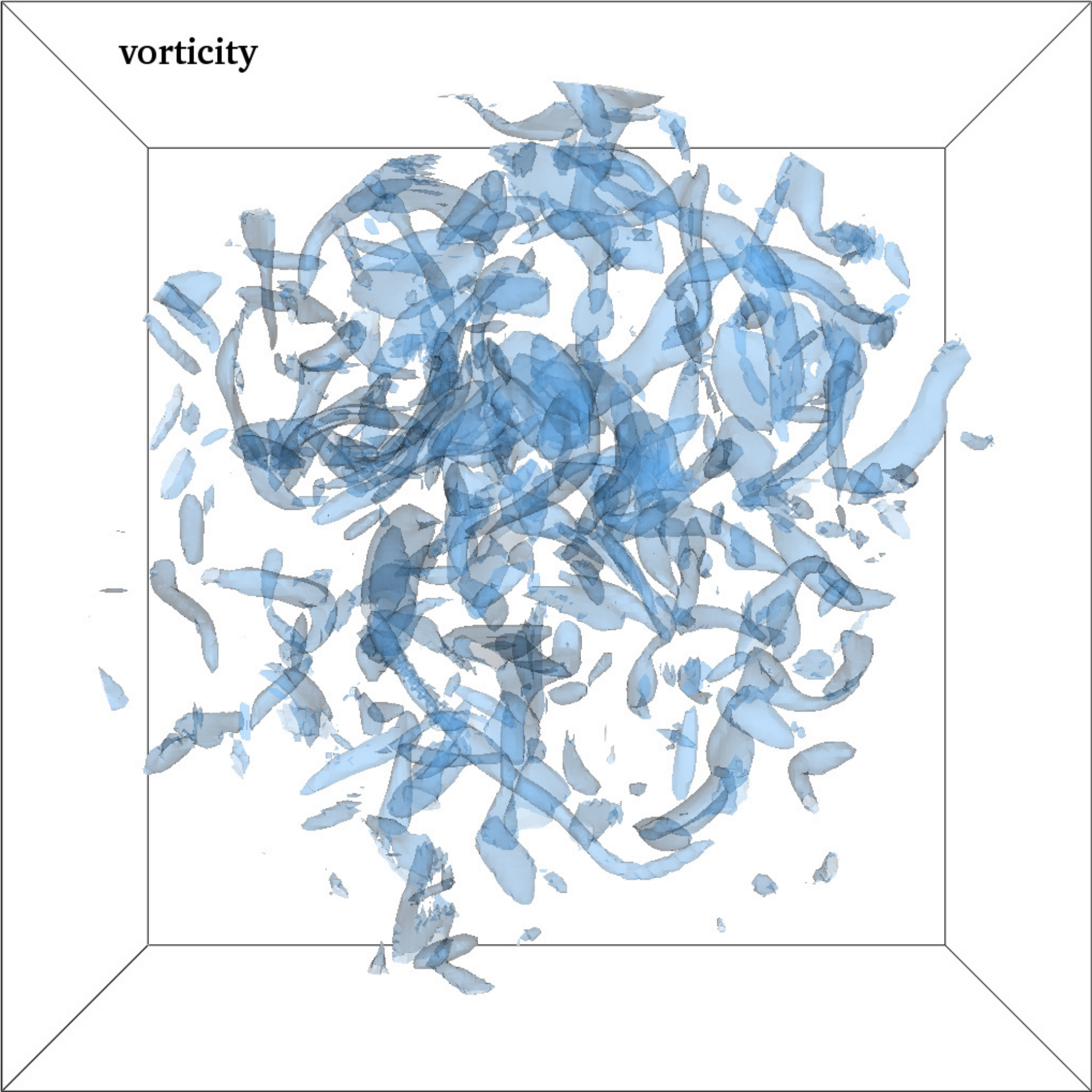} &
\includegraphics[width=0.4\linewidth]{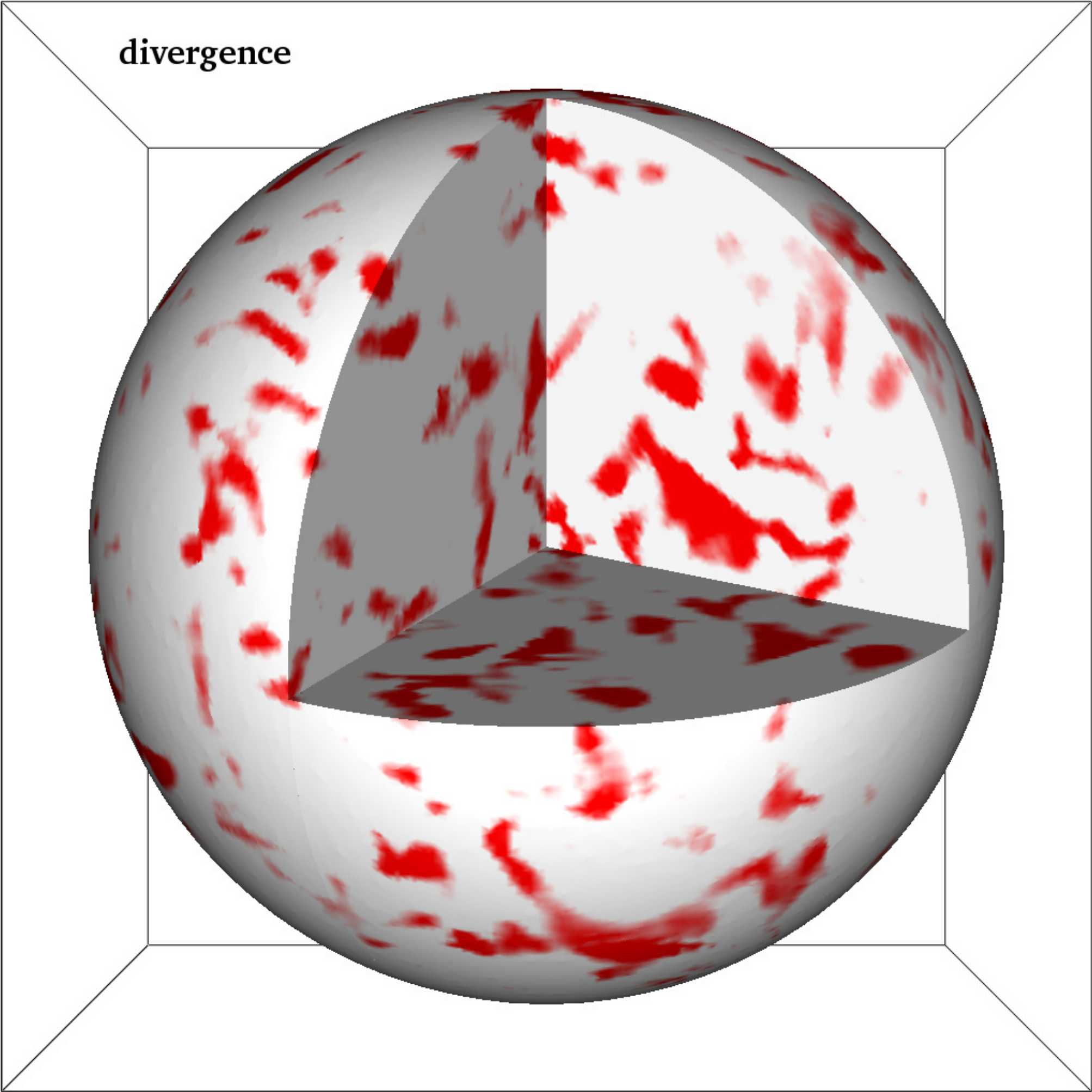}
\end{tabular}
\end{center}
\caption{a) Spherical slice of the gas density inside the Jeans volume at $\tau=12$ for our run with 128 cells per Jeans length. b) Velocity streamlines on a linear color scale ranging from dark blue ($0\,\kms$) to light gray ($5\,\kms$). c) Magnetic field lines, showing a highly tangled and twisted magnetic field structure typical of the small-scale dynamo; yellow:~$0.5\,\mathrm{mG}$, red:~$1\,\mathrm{mG}$. d) Four randomly chosen, individual field lines. The green one, in particular, is extremely tangled close to the center of the Jeans volume. e) Contours of the vorticity modulus, $\left|\nabla\times\vect{v}\right|$, showing elongated, filamentary structures typically seen in subsonic turbulence \citep[e.g.,][]{Frisch1995}. f) Spherical slice of the divergence of the velocity field, $\nabla\cdot\vect{v}$; white:~compression, red:~expansion.}
\label{fig:snapshots}
\end{figure*}

To visualize the structure of the density, magnetic field and turbulence, we show three-dimensional renderings of the density, velocity, magnetic field lines (volume-filling and individual ones), the vorticity, $\left|\nabla\times\vect{v}\right|$, and the divergence of the velocity, $\nabla\cdot\vect{v}$, in Figure~\ref{fig:snapshots}. The central Jeans volume is shown at $\tau=12$, i.e., far into the collapse regime, when --through dynamo action and compression-- the magnetic field has increased by six orders of magnitude to approximately $1\,\mathrm{mG}$. Density fluctuations inside the Jeans volume (upper left panel) are rather weak (see also the radial profiles in Fig.~\ref{fig:radprofiles}) and are mostly correlated with the divergence of the velocity field (lower right). Since the surroundings of the core are subject to contraction, the velocity field is dominated by regions of compression. Only some patchy, cloud-like regions show expansion. The velocity field (upper right) is turbulent, with the outer regions showing some traces of inflow toward the center of the Jeans volume (see also the radial velocity profiles of Fig.~\ref{fig:radprofiles}). The vorticity contours (lower left) are elongated filaments. Some of these filaments seem to extend through the whole Jeans volume. They are folded and twisted several times, similar to the magnetic field lines (middle panels). The magnetic field is extremely tangled, which is typical of the small-scale dynamo. Since the magnetic field is amplified most efficiently by winding-up and folding of the field lines on scales of the smallest resolvable vortices, the field is most strongly twisted on these smallest scales. In the next section, we investigate the scale-dependence of the magnetic field quantitatively by means of magnetic and turbulent energy spectra.

\subsection{Magnetic field spectra inside the core}
In a proof-of-concept study we suggested in paper I that a small seed magnetic field can be amplified significantly by the small-scale dynamo during the collapse of a dense gas cloud. Here, we provide a more quantitative analysis by investigating Fourier spectra to study the scale-dependence of the turbulence and magnetic field. First, we restrict the analysis to Fourier spectra computed in the collapsing frame of reference. We therefore extract a Cartesian box centered on the maximum density in the core with side length equal to the Jeans length at each time frame (see section~\ref{sec:methods_fourier}).

\begin{figure}[t]
\centerline{\includegraphics[width=1.0\linewidth]{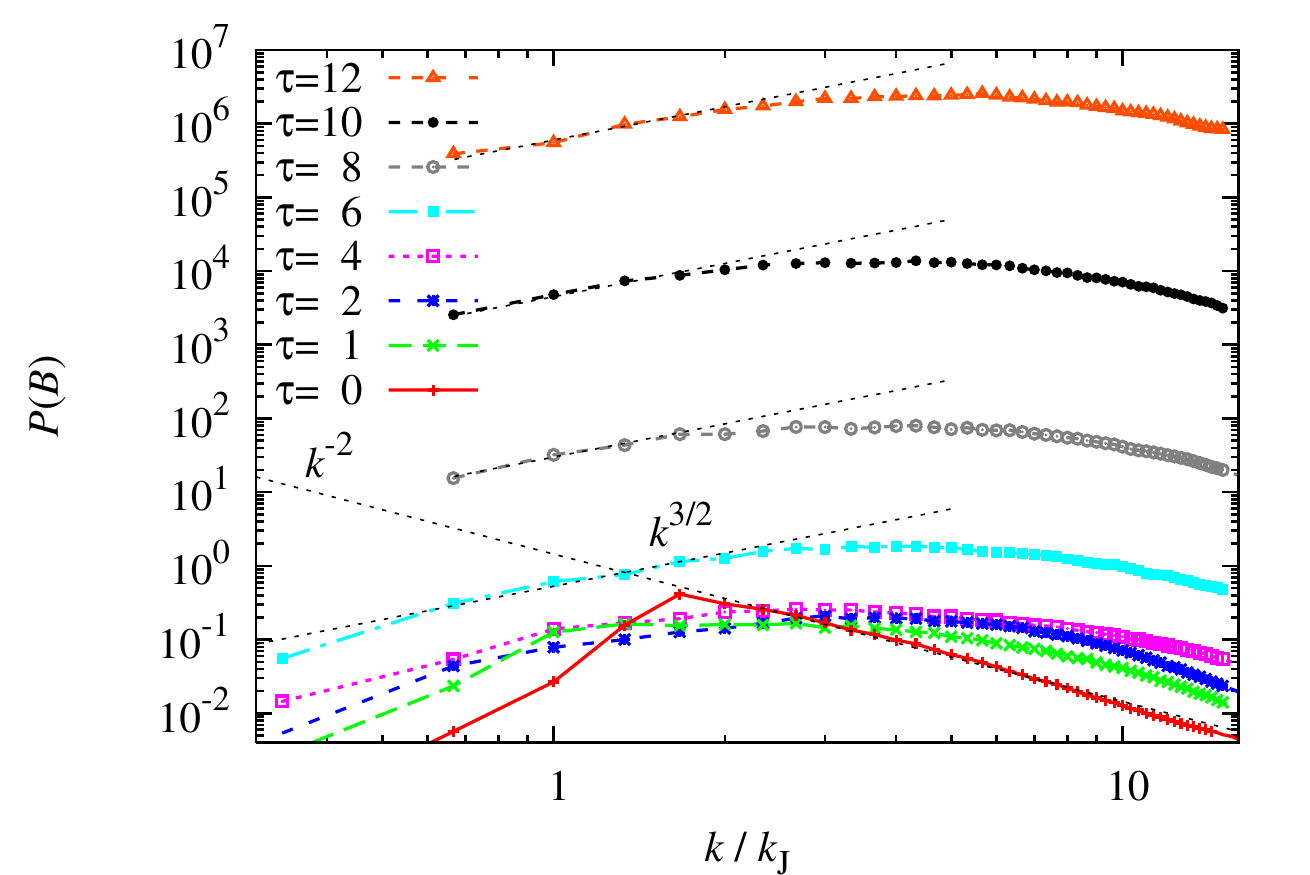}}
\caption{Time evolution of the magnetic field spectra for the run with 128 cells per Jeans length as a function of wavenumber, normalized to the local Jeans wavenumber, i.e., the magnetic field inside the dense core is shown for each timestep, in the collapsing frame of reference. The initial magnetic field spectrum ($\tau=0$) follows the initial power law, $P(B)\propto k^{-2}$ on scales smaller than the peak scale at $k/\kJ\approx1.4$. Within the regime of turbulent decay ($0<\tau<4$), the peak shifts to $k/\kJ\approx5$--6 due to dynamo amplification, which is most efficient on small scales (discussed further in section~\ref{sec:magresol}). The magnetic field spectra are consistent with the generic $k^{3/2}$ Kazantsev spectrum \citep{Kazantsev1968,BrandenburgSubramanian2005} on scales smaller than the Jeans length and larger than the peak. Within the collapse regime, the magnetic energy grows exponentially on all scales within the core.}
\label{fig:spect_jeans}
\end{figure}

Figure~\ref{fig:spect_jeans} shows the magnetic energy spectra as a function of wavenumber $k/\kJ$, i.e., normalized to the local Jeans wavenumber, $\kJ$, for our highest-resolution run (128 cells per Jeans length). We actually used a three times larger extraction volume to include at least some information from outside the core, such that all spectra were computed on a uniform grid with $384^3$ grid cells. We did not extract a larger volume, because the data are given on an adaptively refined grid, which only contains the fluid variables resolved to the highest level inside the Jeans volume. Interpolating the simulation data to significantly larger grids would introduce artifacts. The chosen volume for extraction onto the highest resolution is thus a reasonable compromise to obtain reliable spatial information for both the interior of the core and its closest surroundings.

An important consistency check for the extraction procedure and spectral method is to investigate the initial conditions first. Figure~\ref{fig:spect_jeans} includes the magnetic energy spectrum at $\tau=0$, which shows that our initial power-law dependence, $P(B)\propto k^{-2}$ is reproduced on scales below the peak. Slight deviations from this power law are due to the extraction procedure and Fourier analysis of a non-periodic volume (see section~\ref{sec:methods_fourier}). Strictly speaking, the discrete Fourier analysis implies a periodic dataset, however, the deviations introduced by simply ignoring this are negligible for the present dataset. In addition to the power-law scaling, also the peak position must be reproduced, which we set to $0.8\,\pc$ in the initial conditions (see paper I). With the initial Jeans length of $1.15\,\pc$, this leads to an expected peak position of $k/\kJ=1.4$, in very good agreement with the peak position measured in the spectrum at $\tau=0$.

The peak of the initial magnetic spectrum at about $k/\kJ=1.4$ quickly shifts to smaller scales and stays roughly constant at $k/\kJ\approx4$--6 for $\tau\gtrsim4$. The initial shift of the peak within $\tau\lesssim2$ means that the magnetic field grows faster on smaller scales, as expected from the small-scale dynamo theory. However, the peak does not shift further to smaller scales than $k/\kJ\approx4$--6, which corresponds to about 21--32 grid cells, because vortices, which drive the dynamo, are under-resolved on scales smaller than 30 grid cells \citep{FederrathDuvalKlessenSchmidtMacLow2010}. The resolution dependence is discussed in detail in section~\ref{sec:newjeansresol}.

On scales smaller than the Jeans scale and larger than the peak, the magnetic field spectra are consistent with the theoretical prediction of a $k^{3/2}$-Kazantsev spectrum \citep{Kazantsev1968,BrandenburgSubramanian2005} in the kinematic regime (i.e., the regime in which the magnetic energy is much smaller than the kinetic energy). Indications of the Kazantsev spectrum are also found in simulations of the intra-galaxy cluster medium by \citet{XuEtAl2009,XuEtAl2010}. However, in both their simulations and in ours, the scaling range for measuring the slope is too narrow to draw final conclusions. It is not clear whether the $k^{3/2}$-power law would persist at higher resolution. In turbulence-in-a-box simulations with external forcing by \citet{HaugenBrandenburgDobler2004} with up to $512^3$ grid cells, there is strong indication that the magnetic energy spectra indeed converge to the Kazantsev slope with increasing resolution. As in previous turbulence-in-box calculations with external forcing \citep[e.g.,][]{ChoEtAl2009}, we find that the magnetic energy grows exponentially on all scales, which is a typical feature of the small-scale dynamo \citep{BrandenburgSubramanian2005}, but has (to the best of our knowledge) not been shown before in a gravity-driven turbulent gas core. We also measure the exponential growth rates of the magnetic field as a function of Jeans resolution in section~\ref{sec:newjeansresol} below.

\subsection{Magnetic field spectra in the fixed frame of reference}
The magnetic energy spectra in Figure~\ref{fig:spect_jeans} suggest that the spectra fall off more steeply than the $k^{3/2}$ Kazantsev law toward large scales, outside the Jeans volume \citep[see also the spectra in][]{XuEtAl2010}. In order to clarify this and to investigate the field structure outside the Jeans volume, we introduce and apply a new method to infer the spectrum over more than four orders of magnitude in length scales. Since we aim to investigate the field structure in the fixed frame of reference at the largest available scales in the simulation, but at the same time would like to include spectral information on the very smallest scales, we apply a two-step approach. First, we re-normalize the high-resolution spectra obtained inside the Jeans volume (Fig.~\ref{fig:spect_jeans}), by shifting them to the correct position with respect to the fixed frame of reference. However, the spectra at late times during the collapse, obtained with this method, do not contain any information on scales far outside the Jeans volume. Thus, in the second step, we add spectral information on large scales, by gradually extracting larger boxes, centered on the core at a fixed grid resolution. Using this method, we can test the spectral energy scale-by-scale. We call this method `scale-by-scale extraction'.
\begin{figure}[t]
\centerline{\includegraphics[width=1.0\linewidth]{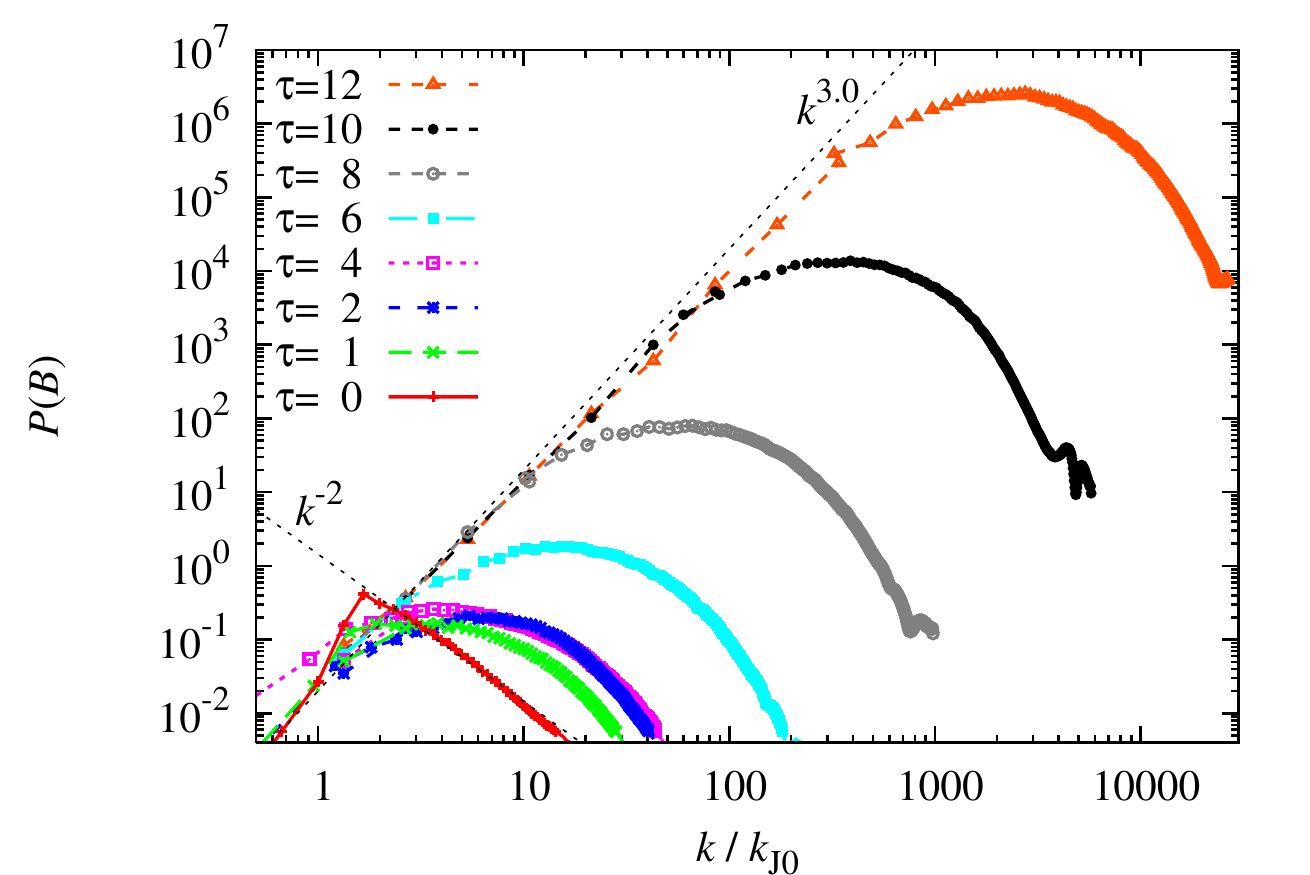}}
\caption{Magnetic energy spectra in the fixed frame of reference, i.e., the spectra are normalized to the initial Jeans wavenumber, $\kJzero$.
The time evolution in the collapse regime shows that the magnetic field is basically frozen-in on large scales, while the turbulent dynamo keeps amplifying the field on smaller and smaller scales as the collapse proceeds.}
\label{fig:spect_fixed}
\end{figure}

The result of this two-step approach is shown in Figure~\ref{fig:spect_fixed}, where we plot the spectra of magnetic energy in the fixed frame of reference, i.e., as a function of the initial (at $\tau=0$) Jeans wavenumber, $\kJzero$, of the collapsing system. The first thing to note is that the spectra on large scales, obtained with the scale-by-scale extraction method connect reasonably well with the spectra computed inside the Jeans volume and re-normalized to match the fixed frame of reference. Some deviation is seen only at $k/\kJzero\approx300$ for the spectrum at $\tau=12$, which can be taken as a measure of the uncertainty in the spectra obtained with our scale-by-scale extraction method. Given the total range of scales that we aim to probe here, the difference in the spectra obtained with our two-step approach is acceptable. We also tested whether extending the scale-by-scale extraction to smaller scales (inside the Jeans volume) matches the high-resolution spectra of step one. We found that they do within an uncertainty of about 25\%, i.e., the slopes and peak positions are reproduced reasonably well with the scale-by-scale extraction method. However, the high-resolution spectra inside the Jeans volume are more accurate, and we thus prefer the two-step approach described above.

Three main results can be extracted from Figure~\ref{fig:spect_fixed}. First, the magnetic field always grows fastest on the smallest available scales in the simulation, with the peak located at around 30 grid cells. Second, as the collapse proceeds, the field is amplified on small scales inside the core, but is essentially frozen-in on large scales. Third, the large-scale spectra follow a power law close to $k^{3.0}$. This particular exponent for the power law is consistent with the radial dependence of the field (see Figs.~\ref{fig:radprofiles} and~\ref{fig:resradprofiles}) and depends strongly on resolution. The magnetic energy on scales outside the core is
\begin{equation}
B_\mathrm{rms}^2 \propto \int P(B)\,dk \propto k^{4.0}\,.
\end{equation}
From this it follows that $B_\mathrm{rms}\propto k^{2.0}\propto r^{-2.0}$ outside the Jeans volume, which is consistent with the power-law behavior of the radial profile of $B_\mathrm{rms}$ in Figure~\ref{fig:radprofiles}. It should be re-emphasized, however, that the exact exponent of this power law is essentially meaningless, as it depends on the numerical Jeans resolution. Higher resolution will lead to a steeper increase of $P(B)$ toward small scales (see Fig.~\ref{fig:resradprofiles}). In reality, the magnetic field will grow much more strongly due to dynamo action than what we can resolve in the present calculation with 128 cells per Jeans length, and thus, our amplification rate is a lower limit (discussed further in section~\ref{sec:newjeansresol}).

\subsection{Probability distribution function of the gas density}
The probability distribution function (PDF) of the gas density is a useful measure of the turbulence in any turbulent system exhibiting density fluctuations. Moreover, the PDF is an essential ingredient for models of star formation \citep[e.g.,][]{PadoanNordlund2002,KrumholzMcKee2005,HennebelleChabrier2008,HennebelleChabrier2009} and the gas distribution in galaxies \citep[e.g.,][]{Tassis2007,KrumholzMcKeeTumlinson2009}. The density PDF has been studied in some detail in non-self-gravitating, turbulent systems \citep[e.g.,][]{Vazquez1994,PadoanNordlundJones1997,PassotVazquez1998,LemasterStone2008,FederrathKlessenSchmidt2008,PriceFederrath2010}, and in turbulent systems including self-gravity \citep[e.g.,][]{Klessen2000,FederrathGloverKlessenSchmidt2008,KainulainenEtAl2009,ChoKim2011, KritsukNormanWagner2011}. However, it has not been analyzed yet in a collapsing system, in which turbulence is replenished by the gravitational collapse of a dense core.

\begin{figure}[t]
\centerline{\includegraphics[width=1.0\linewidth]{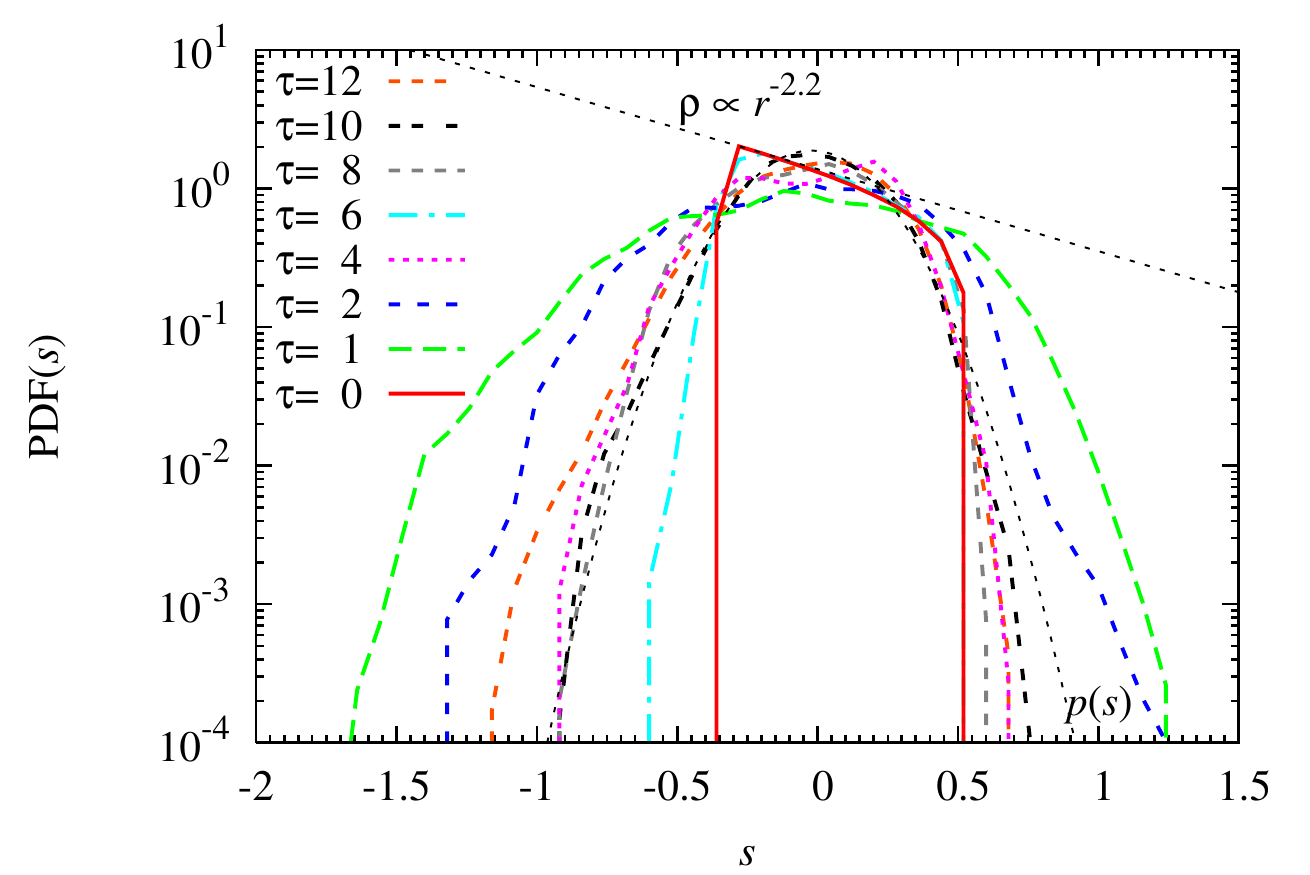}}
\caption{Time evolution of the probability distribution functions (PDFs) of the logarithmic gas density, $s\equiv\ln\left(\rho/\left\langle\rho\right\rangle\right)$, inside the collapsing core. The initial density PDF is characterized by the asymptotic radial profile of a Bonnor-Ebert sphere at low densities, $\rho\propto r^{-2.2}$, while at higher densities, it turns into a steeper PDF due to the flat inner core. The PDFs in the turbulent decay regime ($\tau\lesssim4$) are convolutions of the PDF of the initial density distribution and log-normal distributions typical of turbulence. In the collapse regime ($\tau\gtrsim4$), the PDFs are close to log-normal distributions, with some modification due to the radial density distribution (see Fig.~\ref{fig:radprofiles}). The dotted curve shows a log-normal fit using equation~(\ref{eq:lognormal}) with the imposed condition, eq.~(\ref{eq:pdfmean}) at $\tau=10$.}
\label{fig:pdfs}
\end{figure}

Figure~\ref{fig:pdfs} shows the time evolution of the PDFs of the logarithmic density
\begin{equation}
s\equiv\ln\left( \frac{\rho}{\left\langle\rho\right\rangle}\right)\,,
\end{equation}
where $\left\langle\rho\right\rangle$ denotes the mean density in the core, i.e., inside the Jeans volume. The PDF at $\tau=0$ is purely a result of the initial, radial density distribution, following a Bonnor-Ebert profile. This profile exhibits a flat inner core for radii smaller than the Jeans radius, and can be approximated with a power law of the form $\rho\propto r^{-\alpha}$ with $\alpha\approx2.2$ at large radii \citep{Ebert1955,Bonnor1956}. Using the derived relation between a power-law radial distribution and the corresponding density PDF (see Appendix~\ref{app:pdf}), we can estimate the power-law exponent of the density PDF from the power-law exponent of the radial distribution, $\alpha$. Thus, the PDF of the logarithmic density, $s$, follows a power law for small logarithmic densities with exponent $-3/\alpha$, and falls off more steeply toward higher densities, due to the flattening of the Bonnor-Ebert profile in the center of the core (see Fig.~\ref{fig:radprofiles}).

Both in the regime of turbulent decay ($\tau\lesssim4$), and in the collapse regime ($\tau\gtrsim4$), the volume-weighted PDF develops a log-normal form,
\begin{equation} \label{eq:lognormal}
p_s(s)\,ds=\frac{1}{\sqrt{2\pi\sigma_{s,\mathrm{turb}}^2}}\,\exp\left[-\frac{(s-\langle s \rangle)^2}{2\sigma_{s,\mathrm{turb}}^2}\right]\,ds\,,
\end{equation}
where $\langle s \rangle$ and $\sigma_{s,\mathrm{turb}}$ denote the mean logarithmic density and  standard deviation, respectively. This PDF is typical of compressible, nearly isothermal, turbulent flows, which has been motivated with the central limit theorem \citep{Vazquez1994}. Note that the mean is related to the standard deviation by
\begin{equation} \label{eq:pdfmean}
\langle s \rangle = -\frac{1}{2}\sigma_{s,\mathrm{turb}}^2\,,
\end{equation}
because the mean density inside the Jeans volume must be recovered by integration of the PDF, weighted by gas density, $\langle\rho\rangle=\int \rho\,p_s\,ds$ \citep{Vazquez1994,FederrathKlessenSchmidt2008,FederrathDuvalKlessenSchmidtMacLow2010}. The measured PDFs in Figure~\ref{fig:pdfs}, however, exhibit significant deviations from the log-normal function due to intermittency \citep{KritsukEtAl2007,SchmidtEtAl2009,FederrathDuvalKlessenSchmidtMacLow2010} and the convolution with the radial density distribution inside the Jeans radius of the collapsing Bonnor-Ebert sphere. Note that we do not see the typical power-law PDFs observed in self-gravitating systems \citep[][]{Klessen2000,FederrathGloverKlessenSchmidt2008,KainulainenEtAl2009,ChoKim2011, KritsukNormanWagner2011} during the collapse, because here we always compute PDFs in the collapsing frame of reference, thus ignoring the gas outside the Jeans volume, which has a clear power-law PDF.

In order to compare the PDFs obtained in our collapsing system with results on the PDF reported in the literature, we compute the standard deviations of the PDFs below. The standard deviations of the density PDF are $\sigma_s=0.225$, 0.429, 0.353, and 0.290 for $\tau=0$, 1, 2, and 3 in the turbulent decay regime. It is important to note that the initial dispersion at $\tau=0$ has nothing to do with turbulence. It is purely due to the initial, radial density distribution of the Bonnor-Ebert sphere. In order to separate the standard deviation caused by turbulent compression from the standard deviation due to the radial density distribution inside the Jeans volume, we subtract the initial density dispersion caused by the Bonnor-Ebert profile: $\sigma_{s,\mathrm{turb}}^2 = \sigma_s^2(\tau)-\sigma_s^2(0)$. The standard deviations of the turbulence, $\sigma_{s,\mathrm{turb}}$, can then be compared to analytic estimates based on the rms Mach number, $\mathcal{M}$ \citep{PadoanNordlundJones1997,PassotVazquez1998,FederrathKlessenSchmidt2008,FederrathDuvalKlessenSchmidtMacLow2010,PriceFederrathBrunt2011}:
\begin{equation} \label{eq:sigmaturb}
\sigma_{s,\mathrm{turb}}^2=\ln\left(1+b^2\mathcal{M}^2\right)\,.
\end{equation}
The parameter $b$ in this equation is of order unity and depends on the way of turbulent energy injection \citep{SchmidtEtAl2009}. It can vary by a factor of three, from $b\approx1/3$ for turbulence excited by solenoidal (rotational) modes, to $b\approx1$ for turbulence excited by purely compressive modes \citep{FederrathKlessenSchmidt2008}. Moreover, $b$ increases smoothly from 1/3 to 1, when the mixture of modes for the turbulence driving is smoothly varied from fully solenoidal to fully compressive \citep{FederrathDuvalKlessenSchmidtMacLow2010}. Given the rms Mach numbers of order unity (see paper I), we obtain $b\approx0.42$, 0.40, and 0.33 for $\tau=1$, 2, and 3, respectively, with an uncertainty of $\pm0.05$ each. The values at early times ($\tau=1$, and 2) fit the expected value ($b\approx0.4$) for a natural mixture --which were our initial conditions--  very well \citep{FederrathDuvalKlessenSchmidtMacLow2010}. The further evolution indicates a transition to a more solenoidal behavior of the turbulence. This is reasonable, as $\mathcal{M}$ decreases from unity to about 0.5--0.6 at $\tau=3$, which means that shocks are basically absent at this stage. Some caution should be exercised in deriving $b$ for the subsonic regime, although equation~(\ref{eq:sigmaturb}) seems to hold in both the subsonic and transsonic regimes \citep{PassotVazquez1998}, as well as in the highly supersonic regime of turbulence \citep{PriceFederrathBrunt2011}.

In the collapse regime ($\tau\gtrsim4$), it becomes increasingly difficult to separate the density dispersion caused by the infall profile and caused by the turbulence that is generated during the collapse. The density PDFs are close to log-normal distributions in the collapse regime, with some modifications due to the infall profile. As an example, we show a log-normal fit to the data at $\tau=10$ in Figure~\ref{fig:pdfs} as the dotted line labeled $p(s)$. Using the total standard deviations, $\sigma_s$, and the Mach numbers in the collapse regime, we derive $b\approx0.65$, 0.48, and 0.37 for $\tau=8$, 10, and 12, respectively, again with an uncertainty of about $\pm0.05$. These values are upper limits, because of the additional dispersion from the radial density distribution, which is, however, quite small inside the Jeans volume (see Fig.~\ref{fig:radprofiles}). These results for the collapse regime are indicative of a time evolution of $b$ from a state, when compressive modes are more important (at $\tau\approx8$) to values close to equipartition at later times, which is confirmed with a spectral decomposition into solenoidal and compressible modes of the velocity field in section~\ref{sec:rot_ratio} below.

\subsection{The driving scale and spectrum of gravity-driven turbulence} \label{sec:vel_spect}

As discussed in the introduction, the driving of turbulence by extracting potential energy from the gravitational field of a gas cloud has been suggested in several studies since \citet{Hoyle1953}. But what is the characteristic driving scale of gravity-driven turbulence? Answering this question is important, because the injection scale of turbulence may determine the mode of star formation in molecular clouds, isolated versus clustered \citep{KlessenHeitschMacLow2000,HeitschMacLowKlessen2001,Klessen2001}. Moreover, kinetic energy injected on large scales can produce large-scale coherent gas compressions, promoting the formation of condensations, while turbulence driven on small scales acts to inhibit star formation. Over the last 30 years it has become clear that the turbulence in present-day molecular clouds contains most of the energy on large scales \citep{Larson1981,SolomonEtAl1987,OssenkopfMacLow2002,HeyerBrunt2004}, i.e., it is driven on large scales \citep[][and references therein]{BruntHeyerMacLow2009}. Most important drivers of galactic-scale turbulence are the expansion waves produced by supernova explosions and expanding HII regions around massive star clusters \citep{MacLowKlessen2004,SchneiderEtAl2010}, and the galactic spiral shock, producing large-scale converging flows \citep{Vishniac1994,WalderFolini1996,VazquezSemadeniEtAl2006,HeitschEtAl2006,HennebelleEtAl2008,BanerjeeEtAl2009,FoliniWalderFavre2010}. The power of jets and outflows from low-mass stars is also enormous, and may thus contribute to the turbulence driving on intermediate and small scales \citep[e.g.,][]{TanBlackman2004,NakamuraLi2008,LiEtAl2010,CarrollFrankBlackman2010}.

\begin{figure}[t]
\begin{center}
\def\arraystretch{0.0}
\begin{tabular}{c}
\includegraphics[width=1.0\linewidth]{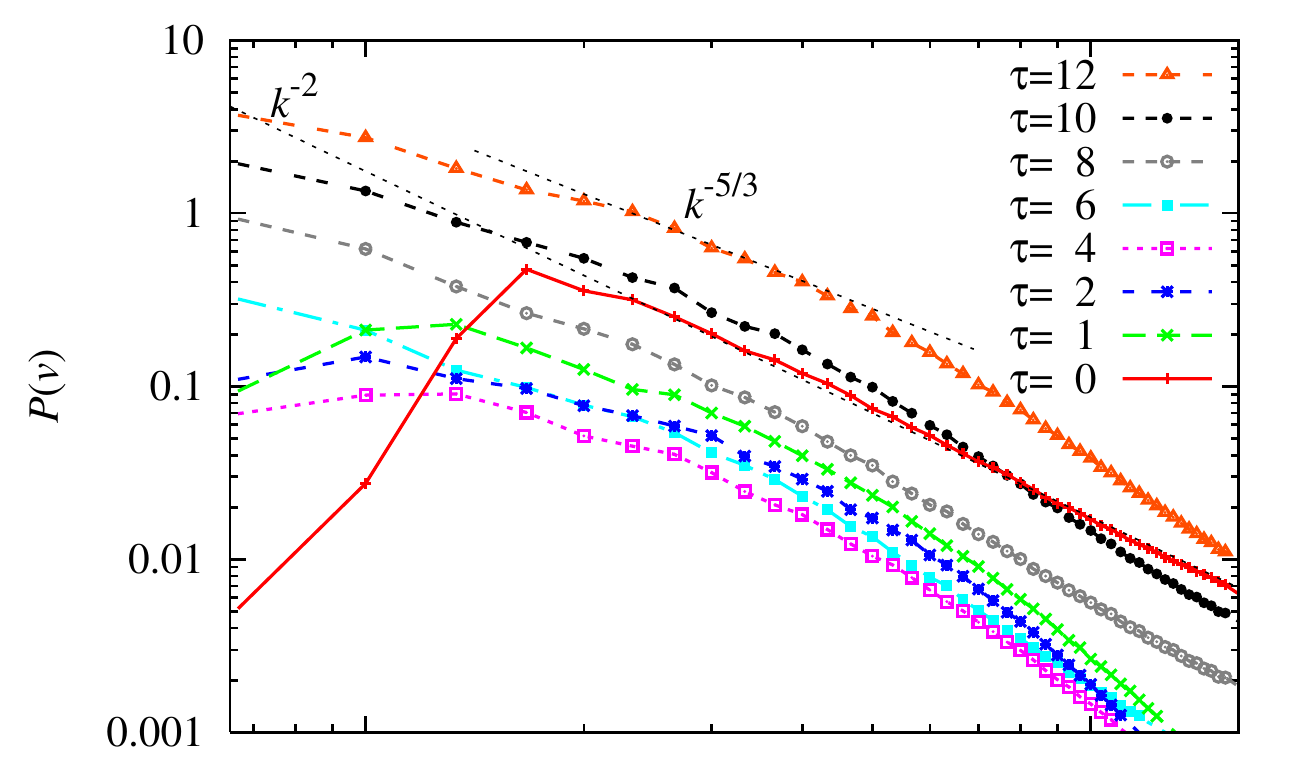} \\
\includegraphics[width=1.0\linewidth]{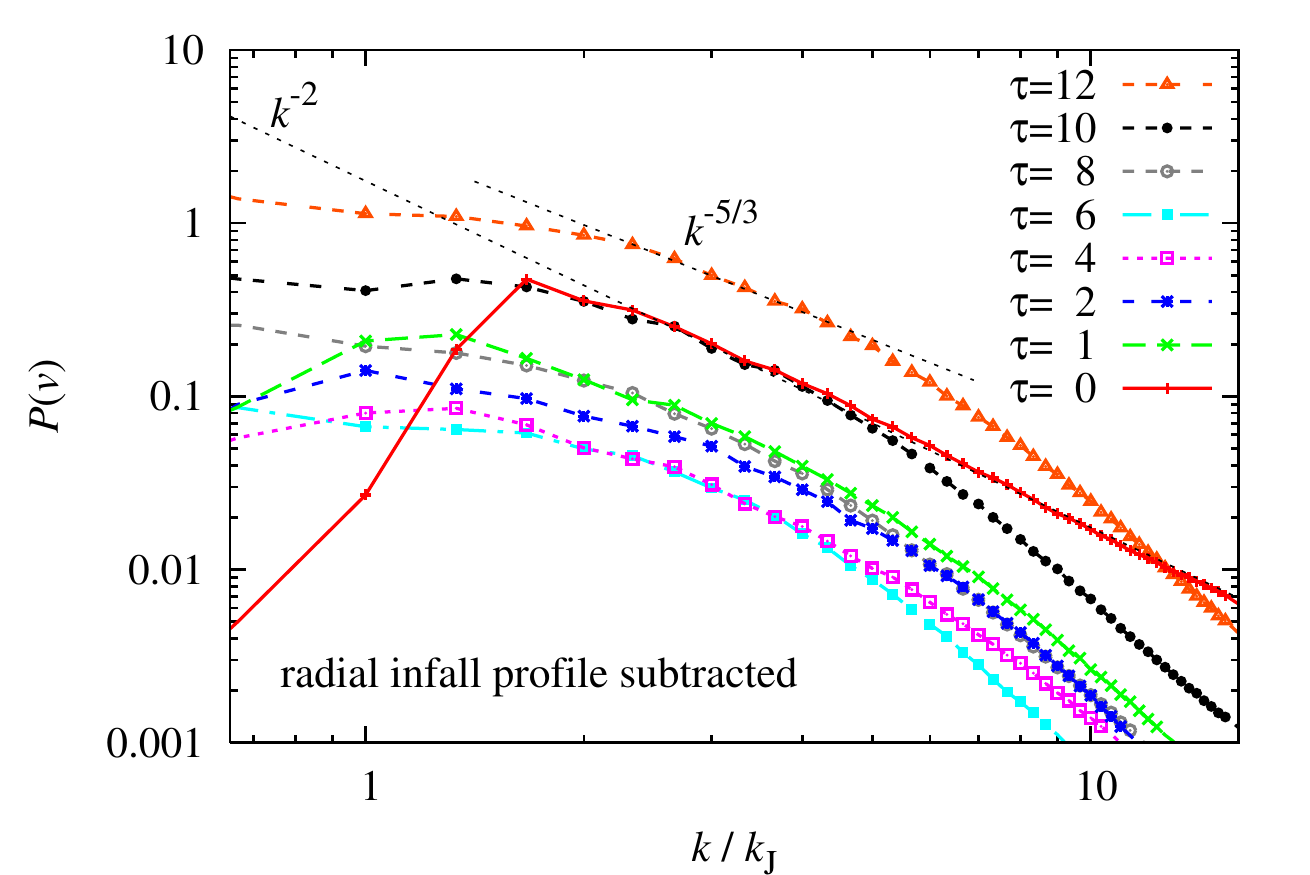}
\end{tabular}
\end{center}
\caption{Top: Velocity power spectra normalized to the local Jeans wavenumber, $k/\kJ$, inside the core for different times until $\tau=12$. Bottom: same as top, but for the infall-subtracted velocity spectra. The flattening of the infall-subtracted spectra around $k/\kJ\approx1$--2 indicates that gravity-driven turbulence exhibits an effective driving scale close to the Jeans scale.}
\label{fig:vel_spect}
\end{figure}

To investigate the driving scale and scale dependence of our gravity-driven turbulent core, we compute velocity spectra inside the Jeans volume of our 128 cell run, defined in analogy to the magnetic field spectra above \citep[e.g.,][]{Frisch1995} as
\begin{equation} \label{eq:vel_spect}
P(v)\,dk = \frac{1}{2}\int\widehat{\vect{v}}\cdot\widehat{\vect{v}}^{*}\,4\pi k^2\,dk\,,
\end{equation}
where $\widehat{\vect{v}}$ denotes the Fourier transform of the velocity field. The total Fourier spectrum can be separated into transverse ($\vect{k} \perp \widehat{\vect{v}}$) and longitudinal ($\vect{k} \parallel \widehat{\vect{v}}$) parts by applying a Helmholtz decomposition. Note that integrating the transverse velocity spectrum yields the specific kinetic energy in transverse (rotational) modes, while integrating the longitudinal velocity spectrum yields the specific kinetic energy in longitudinal (compressible) modes. We will use this decomposition in section~\ref{sec:rot_ratio} to investigate the relative mixture of modes in the turbulence spectrum. First however, we concentrate on the total (transverse + longitudinal) velocity spectrum.

Figure~\ref{fig:vel_spect} shows the velocity Fourier spectra, defined in equation~(\ref{eq:vel_spect}). In the upper panel we show the spectra for the full velocity field, while the bottom panel shows the spectra, where we subtracted the radial infall profile (see Fig.~\ref{fig:radprofiles}) from the velocity field before computing the spectra. In this way, we obtain the pure spectrum of turbulence without direct contributions from infall. As for the magnetic field spectra, the initial velocity power spectrum is reproduced with our Fourier analysis. While the turbulence is decaying ($\tau\lesssim4$), the initial spectrum following $k^{-2}$ flattens slightly and turns into the \citet{Kolmogorov1941c} spectrum, $P(k)\propto k^{-5/3}$, consistent with the expectation for subsonic, hydrodynamic turbulence \citep[e.g.,][]{Frisch1995}. For MHD turbulence, a similar scaling exponent is found \citep[e.g.,][]{ChoVishniac2000,ChoEtAl2009}. Since the turbulence inside the Jeans volume remains subsonic for $\tau\lesssim12$ (see paper I), the Kolmogorov scaling persists until the end of the simulation. The scaling range, however, only extends over the interval $2\lesssim k/\kJ\lesssim 5$, because of the rather low resolution inside the Jeans volume (128 cells). Measuring the power-law exponent of the turbulence accurately would require at least 512 cells inside the Jeans volume as expected from pure turbulence simulations in a periodic box \citep{FederrathDuvalKlessenSchmidtMacLow2010,PriceFederrath2010}.

In the collapse regime, the spectra retain their overall shape, but shift upward with increasing time, which means that the specific turbulent energy, and thus the velocity dispersion increase in the collapse regime (see Fig.~\ref{fig:radprofiles}). The infall-subtracted spectra (bottom panel of Fig.~\ref{fig:vel_spect}) become quite flat around the Jeans scale, which indicates that the turbulence inside the core is driven from the outside with an effective driving scale, $\ell_\mathrm{int}$, approximately on the Jeans scale, $\lJ$. Since the Jeans scale continuously decreases during the collapse, gravity-driven turbulence does not have a fixed driving scale.

This confirms the assumption in \citet{SchleicherEtAl2010} that the integral scale of the turbulence is a fraction of the Jeans scale in their model of dynamo-driven magnetic field amplification during the formation of the first stars and galaxies. They assumed $\ell_\mathrm{int}=0.1\lJ$, while we find in the present numerical experiment that the effective turbulence driving scale is somewhat closer to the Jeans scale, $\ell_\mathrm{int}\lesssim\lJ$.

\section{A new Jeans resolution criterion for (M)HD simulations of self-gravitating gas}
\label{sec:newjeansresol}

In this section, we analyze the resolution dependence of the turbulence and magnetic field growth inside the Jeans volume of our collapsing, magnetized core in more detail. We show that a Jeans resolution of 16 cells is clearly insufficient to obtain dynamo amplification of the magnetic field. Only with a Jeans resolution of 32 cells, we see dynamo amplification, which we explain with the fact that rotational motions of the turbulence (e.g., vortices) are not well resolved, if the Jeans length is resolved with 16 cells or less. In contrast, if the Jeans length is resolved with 32 cells and more, the turbulent energy in the core is converged with resolution. We thus conclude that a minimum Jeans resolution of about 30 grid cells \citep[depending on the particular numerical scheme, see,][]{KitsionasEtAl2009} is required to sufficiently resolve the turbulent energy, and thus the turbulent pressure, and to obtain minimum turbulent dynamo action close to the resolution limit. We emphasize that --apart from a few exceptions-- almost all present-day simulations of self-gravitating magnetized and non-magnetized gas use a Jeans resolution criterion significantly below our threshold value of 30 cells \citep[e.g.,][]{TrueloveEtAl1997}.

\subsection{Resolution dependence of the turbulent dynamo} \label{sec:magresol}

\begin{figure}[t]
\centerline{\includegraphics[width=1.0\linewidth]{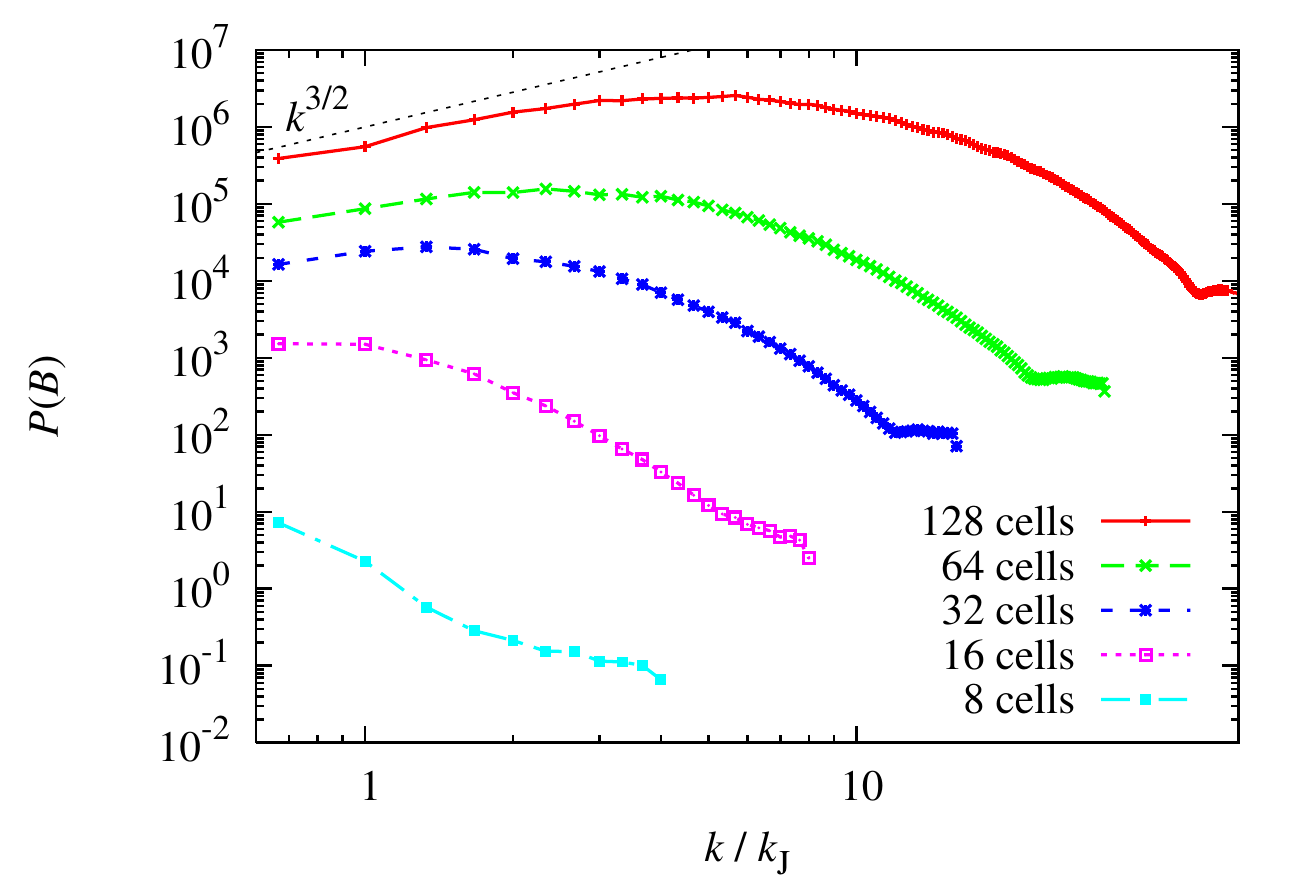}}
\caption{Shows the resolution dependence of the magnetic field spectra at $\tau=12$. The peak of the magnetic energy shifts to smaller scales for increasing Jeans resolution (32, 64, and 128 cells), but is completely absent for a Jeans resolution of 16 cells and below. We conclude that turbulent dynamo amplification of the magnetic field is only obtained, if the Jeans length is resolved with at least 30 grid cells.}
\label{fig:spect_resol}
\end{figure}

The time evolution of the rms magnetic field in paper I and the radial magnetic field profiles of Figure~\ref{fig:resradprofiles} demonstrate a strong resolution dependence of the turbulent dynamo. In this section, we explain the resolution threshold for dynamo action, which we find is about 30 grid cells per Jeans length (see also paper I). Figure~\ref{fig:spect_resol} shows the resolution dependence of the magnetic field spectra at $\tau=12$. Our five runs, in which we resolved the Jeans length with 8, 16, 32, 64, and 128 grid cells are compared. The runs with 32, 64, and 128 cells show a clear maximum of the magnetic energy density on scales below the Jeans scale, i.e., inside the core region, while the 8 and 16 cell runs do not exhibit such a peak. The peak shifts to smaller scales with increasing Jeans resolution: it is located at $k/\kJ\approx1$--2 for 32 cells, $k/\kJ\approx2$--4 for 64 cells, and $k/\kJ\approx4$--6 for 128 cells. The peak in $P(B)$ thus always appears close to the resolution limit, on scales of about 20--30 grid cells, which means that the dynamo amplification of the magnetic field is strongest on these scales. This result, taken together with the absence of a peak for the run with Jeans resolution of 8 and 16 cells suggests that a Jeans resolution of at least 30 grid cells is required for the dynamo to work in a grid simulation. This is significantly more than the resolution criterion of 4 grid cells per Jeans length to avoid artificial fragmentation found by \citet{TrueloveEtAl1997} and frequently applied in simulations involving self-gravity with grid codes \citep[e.g.,][]{KrumholzMcKeeKlein2004} and correspondingly with particle codes \citep[e.g.,][and references therein]{BateBonnellPrice1995,PriceBate2007,FederrathBanerjeeClarkKlessen2010}. The explanation for this is that magnetic field amplification by the turbulent dynamo is most efficient on the smallest scales due to the fast eddy turnover times on the smallest scales \citep[e.g.,][for a review on turbulent dynamo amplification]{BrandenburgSubramanian2005}. However, the dynamo feeds from the transverse, solenoidal turbulent motions, which --due to the discretization of the fluid variables-- are only resolved, if they have at least 30 grid cells across. This is consistent with the spectral analysis of high-resolution ($1024^3$) fixed-grid simulations of driven, supersonic turbulence in \citet{FederrathDuvalKlessenSchmidtMacLow2010}, where a clear deficit of rotational turbulent energy was detected below 30 grid cells.

\begin{figure}[t]
\centerline{\includegraphics[width=1.0\linewidth]{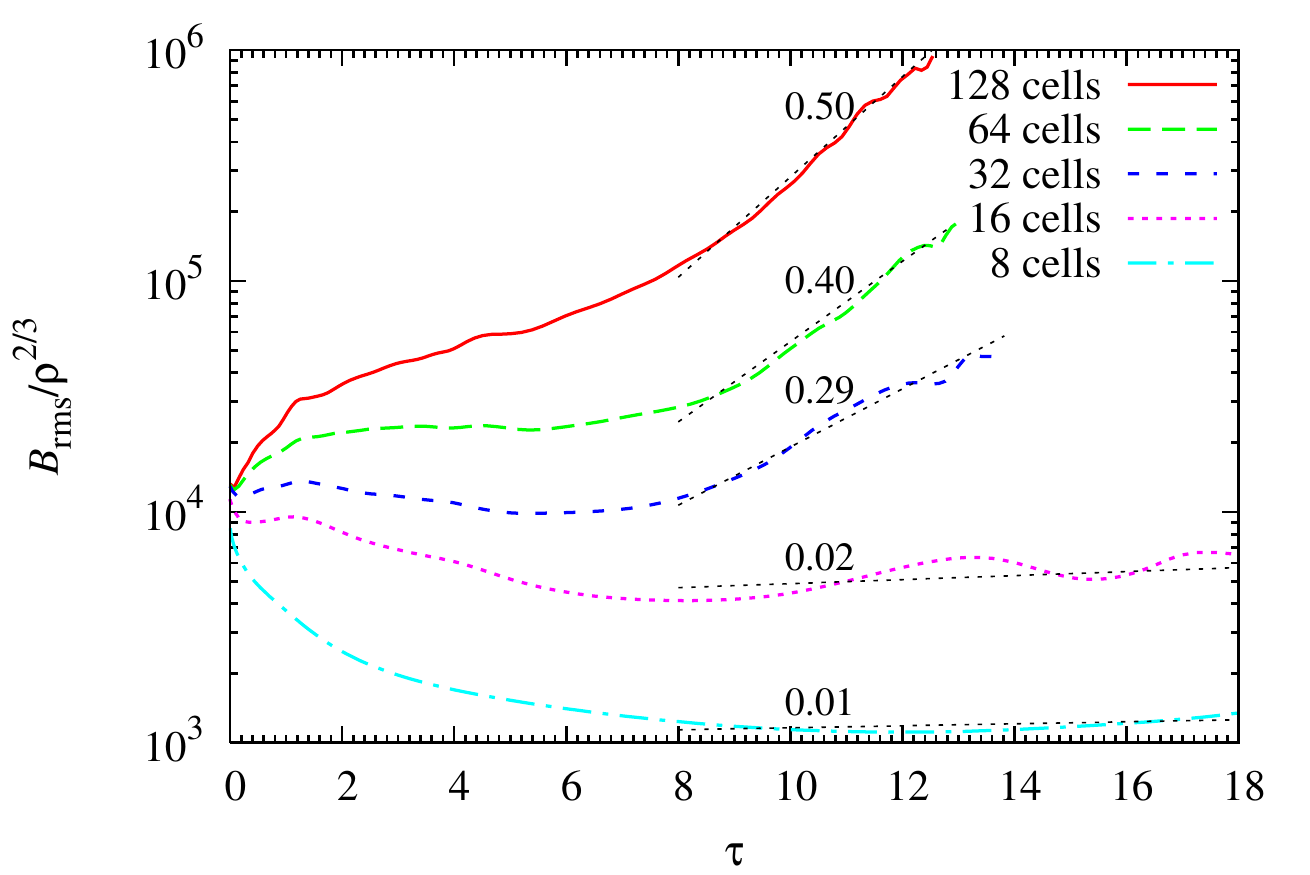}}
\caption{Shows the time evolution of the rms magnetic field strength as a function of Jeans resolution, where we divided out the effect of pure flux freezing, which can provide a maximum possible field amplification by $\rho^{2/3}$ inside the core. The remaining amplification is thus due to the small-scale dynamo. The number on each curve indicates the growth rate, $\Omega$, in the expression $B_\mathrm{rms}/\rho^{2/3}\propto \exp{(\Omega\tau)}$, measured in the interval $\tau=[8,\tau_\mathrm{end}]$.}
\label{fig:time_evol_growth_rates}
\end{figure}

Figure~\ref{fig:time_evol_growth_rates} shows the time evolution of the magnetic field for different Jeans resolutions. The plot shows $B_\mathrm{rms}/\rho^{2/3}$, i.e., we divided out the maximum possible amplification by pure flux-freezing in spherical symmetry. In this representation, we can measure the field growth caused by the turbulent dynamo (see paper I). The growth rate, $\Omega$, in the expression $B_\mathrm{rms}/\rho^{2/3}\propto \exp{(\Omega\tau)}$, is indicated as a label on each curve. For 8 and 16 cells, no clear dynamo amplification occurs, while the growth rate increases discontinuously to $\Omega=0.29$ for a Jeans resolution of 32 cells, marking the onset of dynamo action.

We note that the initial decrease of $B_\mathrm{rms}/\rho^{2/3}$ for the runs with 8 and 16 cells for $\tau\lesssim6$ is mostly due to the turbulent decay and not strongly affected by numerical diffusion of the magnetic field. We estimate the effects of numerical diffusion in Appendix~\ref{app:diffusion}, which may account for a deviation from ideal MHD by $\epsilon<0.005$ in $B\propto\rho^{2/3-\epsilon}$ for any of the runs. The 8 and 16 cell runs show a significantly lower level of turbulence than all the other runs (see paper I, Fig.~3e), because the turbulent energy is not sufficiently resolved in those runs, which is discussed in detail in section~\ref{sec:rot_ratio}.

\begin{figure}[t]
\centerline{\includegraphics[width=1.0\linewidth]{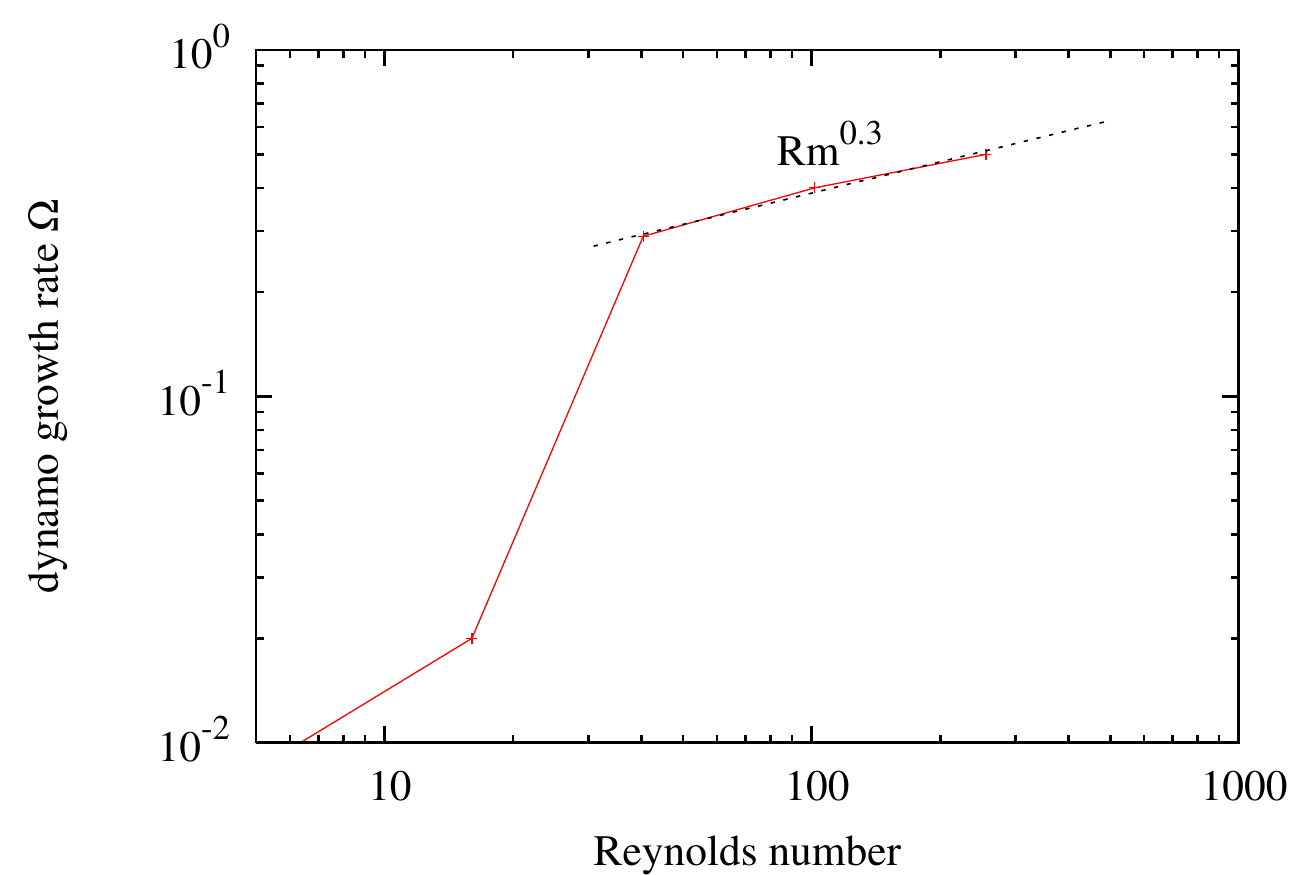}}
\caption{Dynamo growth rate, $\Omega$, measured in Fig.~\ref{fig:time_evol_growth_rates} as a function of Reynolds number. Note the strong increase in the growth rate for $\mathrm{Rm}\lesssim40$, which corresponds to a Jeans resolution of about 30 grid cells.}
\label{fig:growth_rates}
\end{figure}

To see the discontinuous increase of the growth rates with threshold resolution of about 30 cells more clearly, we plot the growth rates as a function of Reynolds number in Figure~\ref{fig:growth_rates}. For computing the Reynolds number, we assumed that the most dissipative wavenumber in the simulation, $k_\eta=N/2$, corresponding to two grid cells \citep[similar to][]{HaugenBrandenburgDobler2004}:
\begin{equation}
\mathrm{Rm}\equiv\left(\frac{k_\eta}{\kJ}\right)^{4/3} \propto N^{4/3}\,,
\end{equation}
with the number of grid cells per Jeans length $N$. Note the strong increase in the growth rate for $\mathrm{Rm}\lesssim40$ in Figure~\ref{fig:growth_rates}. For $\mathrm{Rm}\gtrsim40$, the increase in the growth rate can be approximated with a power law, $\Omega\propto\mathrm{Rm}^{0.3}$, slightly shallower than the theoretical dependence on Reynolds number ($\Omega\propto\mathrm{Rm}^{0.5}$, see Appendix~\ref{app:growth_rate}). In turbulence-in-a-box studies without gravity, a critical Reynolds number of about 35 was found for the onset of dynamo action \citep{HaugenBrandenburgDobler2004}, in excellent agreement with the present study of a gravity-driven turbulent core.

\subsection{The mixture and resolution dependence of turbulent compressible and solenoidal modes} \label{sec:rot_ratio}

In Figures~\ref{fig:resradprofiles}, \ref{fig:spect_resol}, and~\ref{fig:time_evol_growth_rates} we showed that the small-scale turbulent dynamo only starts to operate, if the Jeans length is resolved with about 30 grid cells. Why is there such a threshold resolution? In this section we show that the solenoidal (rotational) turbulent motions, which drive the dynamo are severely under-resolved, if the Jeans length during the collapse is resolved with 16 cells or less. On the other hand, the solenoidal turbulent energy converges for 32 cells per Jeans length and higher resolution, indicating a threshold resolution between 16 and 32 cells per Jeans length for the dynamo to operate.

\begin{figure}[t]
\centerline{\includegraphics[width=1.0\linewidth]{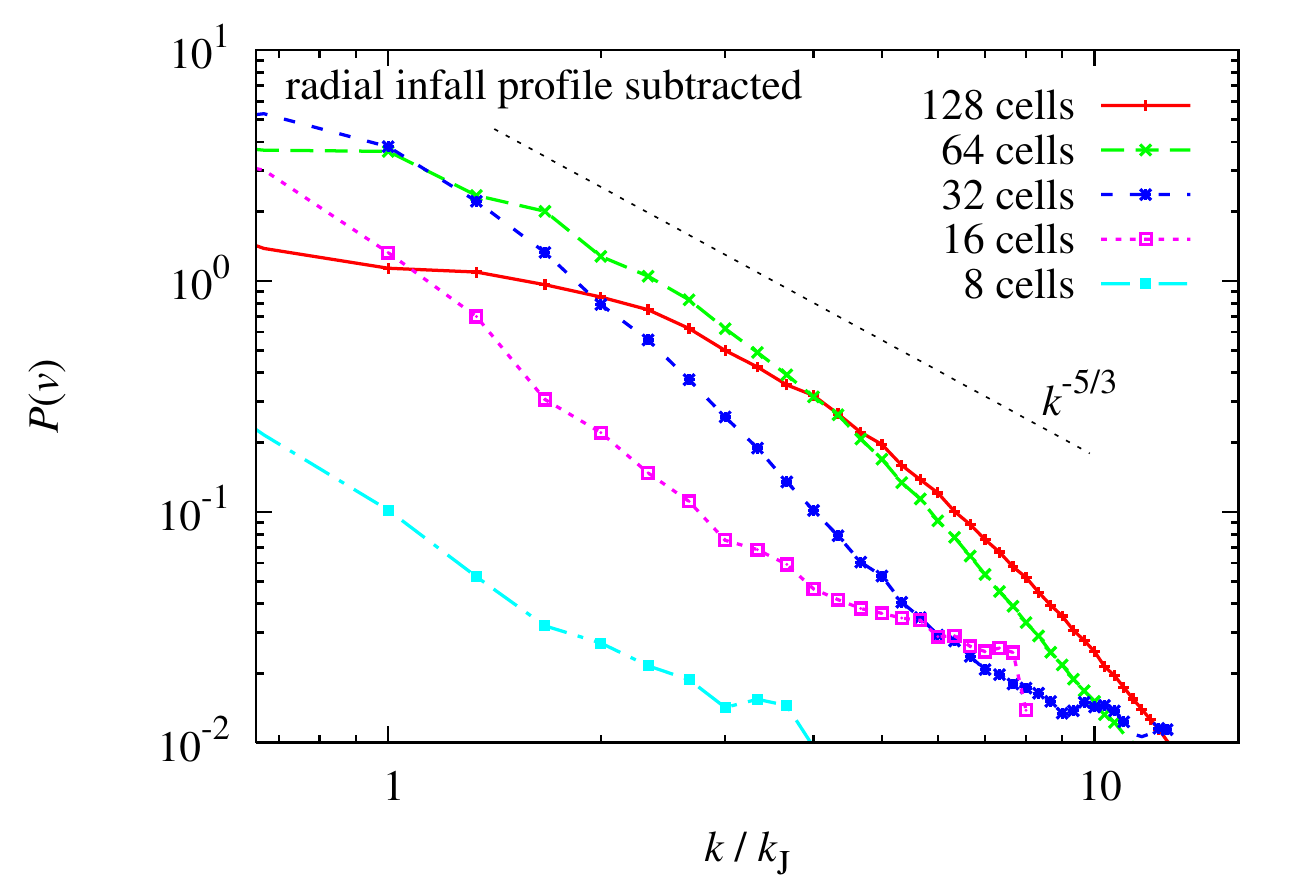}}
\caption{Shows the resolution dependence of the velocity spectra at $\tau=12$. The spectra do not show convergence, even with a Jeans resolution of 128 cells. We expect that a resolution of $512^3$ grid cells inside the Jeans volume is required to obtain an inertial range. The expected \citet{Kolmogorov1941c} scaling of the turbulence in the inertial range is shown only for comparison. The total turbulent energy, i.e., the integral over the spectra, however, indicates convergence for a Jeans resolution of 32 cells and more (see Fig.~\ref{fig:rot_ratio}).}
\label{fig:spect_vel_resol}
\end{figure}

First, we plot the resolution dependence of the infall-subtracted velocity spectra (equation~\ref{eq:vel_spect}) in Figure~\ref{fig:spect_vel_resol}. These spectra are not converged. Although the 128 cell run seems consistent with the expected \citet{Kolmogorov1941c} scaling for subsonic turbulence on some length scales, there is no true inertial scaling range. This is not surprising, given that previous turbulence-in-a-box simulations of driven turbulence showed that inertial range scaling requires at least 512 cells \citep[e.g.,][]{HaugenBrandenburgDobler2004,KritsukEtAl2007,SchmidtEtAl2009,FederrathDuvalKlessenSchmidtMacLow2010}. Although the spectral form is clearly not converged at a Jeans resolution of 128 cells, the integral over the velocity spectrum, i.e., the turbulent energy or turbulent pressure shows indeed convergence for a Jeans resolution of 32 cells and higher.

\begin{figure}[t]
\begin{center}
\def\arraystretch{0.0}
\begin{tabular}{c}
\includegraphics[width=1.0\linewidth]{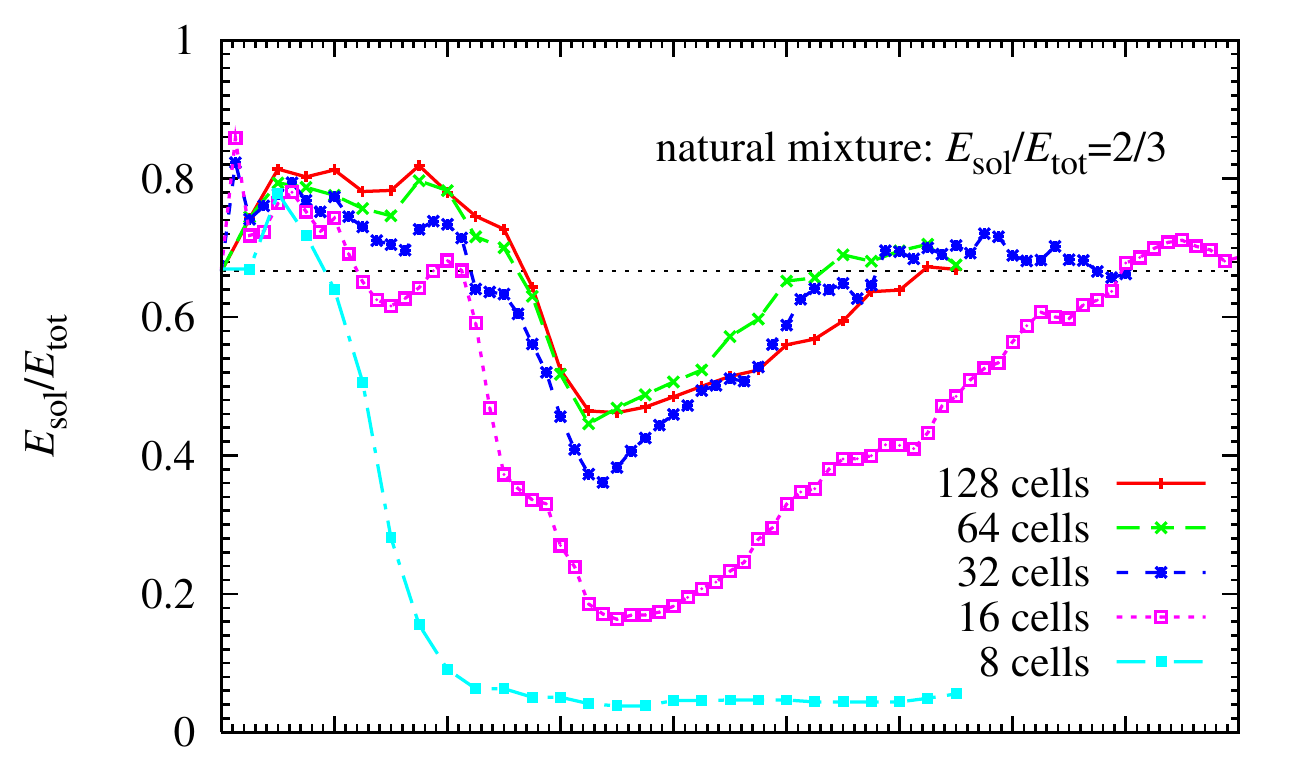} \\
\includegraphics[width=1.0\linewidth]{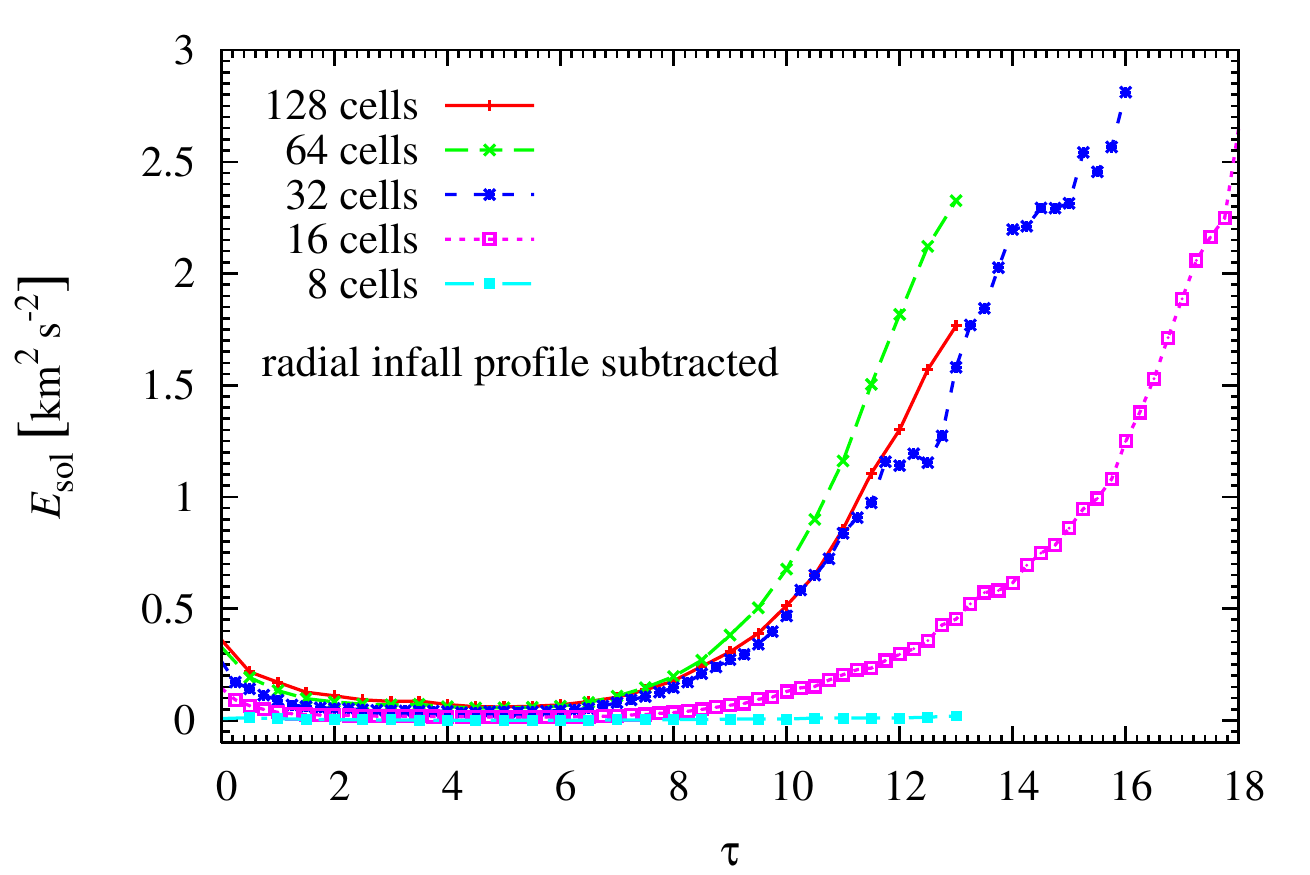}
\end{tabular}
\end{center}
\caption{Top: time evolution of the solenoidal ratio, $\chi=\solratio$, eq.~(\ref{eq:rot_ratio}), for Jeans resolutions of 8, 16, 32, 64, and 128 grid cells. Bottom: Turbulent energy in solenoidal modes for the infall-subtracted motions inside the core. The runs with 32, 64, and 128 cells indicate convergence of $\chi$ (top panel) and in the absolute rotational energy, $E_\mathrm{sol}$ (bottom panel), while the 8 and 16 cell runs clearly underestimate the amount of rotational energy.}
\label{fig:rot_ratio}
\end{figure}

Figure~\ref{fig:rot_ratio} (top panel) shows the solenoidal ratio inside the gravity-driven turbulent core. We define the solenoidal ratio as the specific kinetic energy in rotational turbulent motions divided by the total specific kinetic energy on scales smaller than the Jeans scale,
\begin{equation} \label{eq:rot_ratio}
\chi\equiv\frac{E_\mathrm{sol}}{E_\mathrm{tot}}=\frac{\int_{\kJ}^{\infty}{P_\mathrm{sol}(v)dk}}{\int_{\kJ}^{\infty}{P(v)dk}}\,.
\end{equation}
The solenoidal ratio at $\tau=0$ is $\chi=2/3$, which is the natural mixture \citep{ElmegreenScalo2004,FederrathKlessenSchmidt2008}, imposed as an initial condition. About 20\% of the initial compressible modes are dissipated very quickly, because the turbulence is decaying, and thus $\chi$ increases from 2/3 to about 0.8 in the turbulent decay regime, $\tau\lesssim4$. In the first stages of the collapse regime ($4\lesssim\tau\lesssim7$), $\chi$ drops to about 40--50\%. Within the collapse phase, the kinetic energy provided by the gravitational collapse is gradually converted into rotational motions, which drive dynamo amplification of the magnetic field. For $\tau\gtrsim12$, the solenoidal ratio seems to approach 2/3, which corresponds to the natural ratio in three-dimensional turbulence \citep{FederrathKlessenSchmidt2008}. This natural ratio is most easily pictured by recalling the number of spatial directions in which a longitudinal wave can compress the gas in a three-dimensional system. It is 1 out of 3, which leaves 2/3 for rotational motions.

Gravity-driven turbulence inside galactic disks in two-dimensional (2D), high-resolution simulations by \citet{WadaMeurerNorman2002} show comparable values of $\chi\approx1/2$ to what we find here. In 2D, $\chi$ is indeed expected to be $1/2$ from the dimensional arguments above. However, 2D turbulence is different from three-dimensional (3D) turbulence, as there is an inverse cascade in 2D, i.e., vortices merge to form larger vortices instead of breaking up into smaller vortices as in the 3D case. Thus, the turbulent dynamo is expected to behave (if at all present) very differently from the 3D case studied here.

The evolution of the solenoidal ratio toward more solenoidal turbulence for $\tau\gtrsim7$ explains that the growth rates in Figure~\ref{fig:time_evol_growth_rates} increase after $\tau=7$, showing a steepening of $B_\mathrm{rms}/\rho^{2/3}$ with time for $\tau\approx7$--11, which correlates with the increase in the solenoidal ratio. This is because the dynamo feeds from the solenoidal modes of the turbulence, which increase between $\tau=7$ and 11. Since the solenoidal ratio varies strongly with time in the early phases of the collapse, we had to choose a rather late time within the collapse regime to measure the growth rates of the magnetic field, i.e., we estimated the growth rates in the interval $\tau=[8,\tau_\mathrm{end}]$ in Figure~\ref{fig:time_evol_growth_rates}. Ideally, we would like to measure the growth rates after the solenoidal ratio has converged to $\chi=2/3$, but due to the computational expense, we could not follow the high resolution simulations (64 and 128 cells) far into this regime. The 32 cell run, however, does show a convergence of $\chi$ toward 2/3.

Both the solenoidal ratio and the absolute specific kinetic energy in rotational motions, $E_\mathrm{sol}$, shown in Figure~\ref{fig:rot_ratio}, indicate convergence for a Jeans resolution of 32 cells and higher. Using 8 and 16 cells to resolve the Jeans length is clearly insufficient to resolve rotational motions. We conclude that at least 30 grid cells per Jeans length must be used in (M)HD simulations of self-gravitating gas to resolve the turbulent energy, and thus the turbulent pressure, and to obtain minimal dynamo amplification of the magnetic field on the Jeans scale.

\section{Conclusions and implications}
We presented high-resolution magnetohydrodynamical simulations of the collapse of a dense, magnetized gas cloud \citep[see also,][paper I]{SurEtAl2010}. During the collapse of the cloud, gravitational, potential energy is converted into turbulent motions, which in turn amplify the magnetic field exponentially fast by the turbulent dynamo process. The exponential amplification is driven by the stretching, twisting, and folding of the small-scale magnetic field lines (Fig.~\ref{fig:snapshots}). At sufficiently high Reynolds numbers, even extremely small initial seeds of the magnetic field are expected to be amplified to dynamically important magnetic field strengths on timescales much shorter than the collapse timescales. We conclude that magnetic fields should not be neglected in both primordial and contemporary studies of star formation.

We investigated the scale-dependence of the turbulence and magnetic field by means of Fourier analysis in the collapsing frame of reference (Fig.~\ref{fig:spect_jeans}), showing some indication of the Kazantsev spectrum of turbulent dynamo amplification, and in the fixed frame of reference over more than four orders of magnitude in spatial scale (Fig.~\ref{fig:spect_fixed}). We find that the effective kinetic energy injection scale of gravity-driven turbulence is close to the Jeans length (Fig.~\ref{fig:vel_spect}).

Our Fourier analysis of the magnetic field shows that the dynamo is only excited, if the Jeans length is sufficiently resolved (Fig.~\ref{fig:spect_resol}). The radial dependence of the magnetic field is significantly steeper than what is expected from purely dragged-in magnetic field lines, i.e., flux-freezing ($B_\mathrm{rms}\propto r^{-4/3}$). For a Jeans resolution of 128 cells, we obtained $B_\mathrm{rms}\propto r^{-2.0}$, which is expected to steepen further with increasing Jeans resolution (Fig.~\ref{fig:resradprofiles}). We find that at least 30 grid cells per Jeans length are required for minimum dynamo action to occur in collapse simulations, a resolution requirement, which is not achieved in most current simulations. Here, we studied dynamo amplification with a resolution up to 128 cells per Jeans length, while usually less than 16 cells are used. The amplification rate increases with resolution (Fig.~\ref{fig:time_evol_growth_rates}), which renders any existing simulation of dynamo amplification of a highly turbulent medium insufficiently resolved to determine the physical growth rate of the magnetic field, and can at best provide lower limits on the physical growth rates.

We find that the probability distribution function of the gas density inside the gravity-driven, turbulent Jeans volume follows a log-normal distribution (Fig.~\ref{fig:pdfs}). The standard deviations and Mach numbers inside the core indicate that gravity-driven turbulence approaches a natural mixture of solenoidal and compressible modes, $\solratio\approx2/3$, after a phase of more compressively driven turbulence, caused by the global collapse of the system (Fig.~\ref{fig:rot_ratio}). The turbulent energy (or turbulent pressure) converges only for a Jeans resolution exceeding 30 grid cells. In contrast, the solenoidal component of the turbulent energy is severely under-estimated, if the Jeans length is resolved with 16 cells or less. Thus, we suggest a new Jeans resolution criterion of 30 grid cells per Jeans length to obtain converged results of the turbulent energy on the Jeans scale, and to capture minimum magnetic field amplification by the turbulent dynamo process.

The importance of magnetic fields in present-day accretion disks is generally accepted. However, since even small initial seeds of the magnetic field are efficiently amplified by turbulent dynamo action, it cannot be excluded that magnetic fields also play an important role in primordial accretion disks. Simulations of primordial gas show that it is highly turbulent \citep[e.g.,][]{AbelBryanNorman2002,OSheaNorman2007,WiseAbel2007,ClarkGloverKlessen2008,GreifEtAl2008}, which suggests that there is sufficient kinetic energy in rotational modes of the turbulence \citep[e.g., vorticity; see in particular,][]{WiseAbel2007,GreifEtAl2008} to drive the small-scale dynamo also in primordial gas clouds and accretion disks \citep[see also,][]{PudritzSilk1989}. 

Observed turbulent Mach numbers in e.g., present-day proto-planetary disks are of the order of 0.1--0.5 \citep[][]{HughesEtAl2011}. It is typically believed that this disk turbulence is driven by the magneto-rotational instability \citep[MRI,][]{BalbusHawley1991}. However, in particular in the early phases of star formation, disk turbulence is possibly driven by the gravitational infall and accretion of gas from the envelope onto the disk \citep{KlessenHennebelle2010}. The interaction of the MRI with self-gravitational instabilities may effectively decrease the accretion rate and change the disk morphology \citep{FromangEtAl2004}, also in the primordial case \citep{SilkLanger2006}. A spectacular example of dynamically important magnetic fields in the accretion disks of young stars is the generation of high-speed, bipolar jets, which are launched by a tower of magnetic pressure and/or by a centrifugal magnetic field \citep[e.g.,][for theoretical work on jets and outflows]{BlandfordPayne1982,PudritzNorman1983,Contopoulos1995,LyndenBell2003,MachidaEtAl2006,BanerjeePudritz2006,BanerjeePudritz2007,DuffinPudritz2009}\footnote{See also \citet{BeutherEtAl2010} for a recent observational study on \object{IRAS 18089--1732}, in which --as in our simulations-- the turbulent energies are still dominating the magnetic energy inside the dense core.}. However, even before that stage, the turbulent dynamo may have amplified the magnetic field enough that it could reduce or even suppress fragmentation \citep{MachidaEtAl2008,HennebelleTeyssier2008,BuerzleEtAl2010,PetersEtAl2011}, thus potentially influencing the initial mass function of stars in both primordial and contemporary star formation.

\acknowledgements
We thank Sebastien Fromang for helpful discussions on the spectral analysis of non-periodic datasets.
C.F.~has received funding from the European Research Council under the European Community's Seventh Framework Programme (FP7/2007-2013 Grant Agreement no.~247060) for the research presented in this work. C.F.,~R.B.,~and R.S.K.~acknowledge subsidies from the Baden-W\"urttemberg-Stiftung (grant P-LS-SPII/18) and from the German Bundesministerium f\"ur Bildung und Forschung via the ASTRONET project STAR FORMAT (grant 05A09VHA).
S.S.~thanks the German Science Foundation (DFG) for financial support via the priority program 1177 ``Witnesses of Cosmic History: Formation and Evolution of Black Holes, Galaxies and their Environment'' (grant KL 1358/10).
D.S.~thanks for funding from the European Community's Seventh Framework Programme (FP7/2007-2013) under grant agreement No~229517.
R.B. is funded by the Emmy-Noether grant (DFG) BA 3706.
Supercomputing time at the Leibniz Rechenzentrum (projects pr32lo and h1221) and the Forschungszentrum J\"ulich (projects hhd20 and hhd14) are gratefully acknowledged.
The software used in this work was in part developed by the DOE-supported ASC / Alliance Center for Astrophysical Thermonuclear Flashes at the University of Chicago. Figure~\ref{fig:snapshots} was produced with the open-source visualization software \textsc{visit}.


\begin{appendix}

\section{Power-law density probability distribution functions} \label{app:pdf}
Given a power-law, spherical density distribution with exponent $\alpha$,
\begin{equation}
\rho(r) = \rho_0 \left(\frac{r}{r_0}\right)^{-\alpha} \;.
\end{equation}
we can derive the power-law exponent of the probability distribution function (PDF) of the density. The differential volume of the spherical gas distribution is a shell with
\begin{equation}
dV = 4\pi r^2 dr \;.
\end{equation}
The volume-weighted PDF of density, $p_\rho$ is given by
\begin{equation}
p_\rho \propto \frac{dV}{d\rho} = \frac{dV}{dr}\frac{dr}{d\rho} \propto \left(\frac{\rho}{\rho_0}\right)^{-\left(\frac{3}{\alpha}+1\right)} \;.
\end{equation}
The PDF, $p_s$, of the \emph{logarithmic} density, $s\equiv\ln\left(\rho/\langle\rho\rangle\right)$, is related to the PDF, $p_\rho$, of the \emph{linear} density \citep[see, e.g.,][]{LiKlessenMacLow2003,FederrathKlessenSchmidt2008} by 
\begin{equation}
p_s = \rho\,p_\rho \;,
\end{equation}
such that
\begin{equation}
p_s \propto \left(\frac{\rho}{\rho_0}\right)^{-\frac{3}{\alpha}} \propto \exp{\left(-\frac{3s}{\alpha}\right)} \;.
\end{equation}

\section{Reynolds number dependence of the dynamo growth rate} \label{app:growth_rate}
In the regime of intermediate to large magnetic Prandtl number, $\mathrm{Pm}=\nu/\eta\gtrsim1$, where $\nu$ is the kinematic viscosity and $\eta$ is the magnetic diffusivity, the dynamo growth rate, $\Omega$ is determined by the vorticity, $\omega=\left|\nabla\times\vect{u}\right|$ on small scales. We thus make the ansatz
\begin{equation} \label{eq:growth_rate_ansatz}
\Omega=a\,\omega\,,
\end{equation}
with a numerical constant $a$ of order unity, which is determined by fitting the growth rate in a time evolution plot of the magnetic field, from numerical experiments (see, e.g., Fig.~\ref{fig:time_evol_growth_rates}). From $\nu\omega^2=b^2\,\sigma_v^3/\ell_\mathrm{int}$ with another dimensionless constant $b$, the integral scale $\ell_\mathrm{int}$ of the turbulence and the velocity dispersion $\sigma_v$ on that scale, and equation~(\ref{eq:growth_rate_ansatz}), it follows that the growth rate
\begin{equation}
\Omega = a\,\sqrt{\frac{b^2\sigma_v^3}{\nu\ell_\mathrm{int}}}=ab\,\frac{\sigma_v}{\ell_\mathrm{int}}\sqrt{\mathrm{Re}}=\frac{ab}{2\ted}\sqrt{\mathrm{Re}}\,,
\end{equation}
with the eddy turnover time $\ted\equiv\ell_\mathrm{int}/2\sigma_v$. This shows that the growth rate for dynamo amplification is expected to increase with the square root of the Reynolds number $\mathrm{Re}=\sigma_v\ell_\mathrm{int}/\nu$. For magnetic Prandtl number, $\mathrm{Pm}=1$, the kinematic and magnetic Reynolds numbers are the same: $\mathrm{Re}=\mathrm{Rm}$, such that $\Omega\propto\mathrm{Rm}^{1/2}$ \citep[see also,][]{HaugenBrandenburgDobler2004}. Since both $\nu$ and $\eta$ are controlled by the numerical cutoff scale in our simulations (see section~\ref{sec:resolutionstudy}), $\mathrm{Pm}\approx1$.

\section{Test on the numerical diffusion of the HLL3R ideal MHD scheme} \label{app:diffusion}
Here we test the numerical diffusion of the magnetic field in the new HLL3R scheme for ideal MHD \citep{WaaganFederrathKlingenberg2011}, implemented in the FLASH code. We use the same setup as in section~\ref{sec:methods}, but set all initial velocities to zero. Unlike the turbulent magnetic field structure in the production runs, we start with a uniform magnetic field in $z$-direction with $B=10^{-9}\,\G$. The magnetic field is dragged in and develops an hour-glass-shaped magnetic field. 

\begin{figure}[t]
\begin{center}
\def\arraystretch{0.45}
\begin{tabular}{c}
\includegraphics[width=0.5\linewidth]{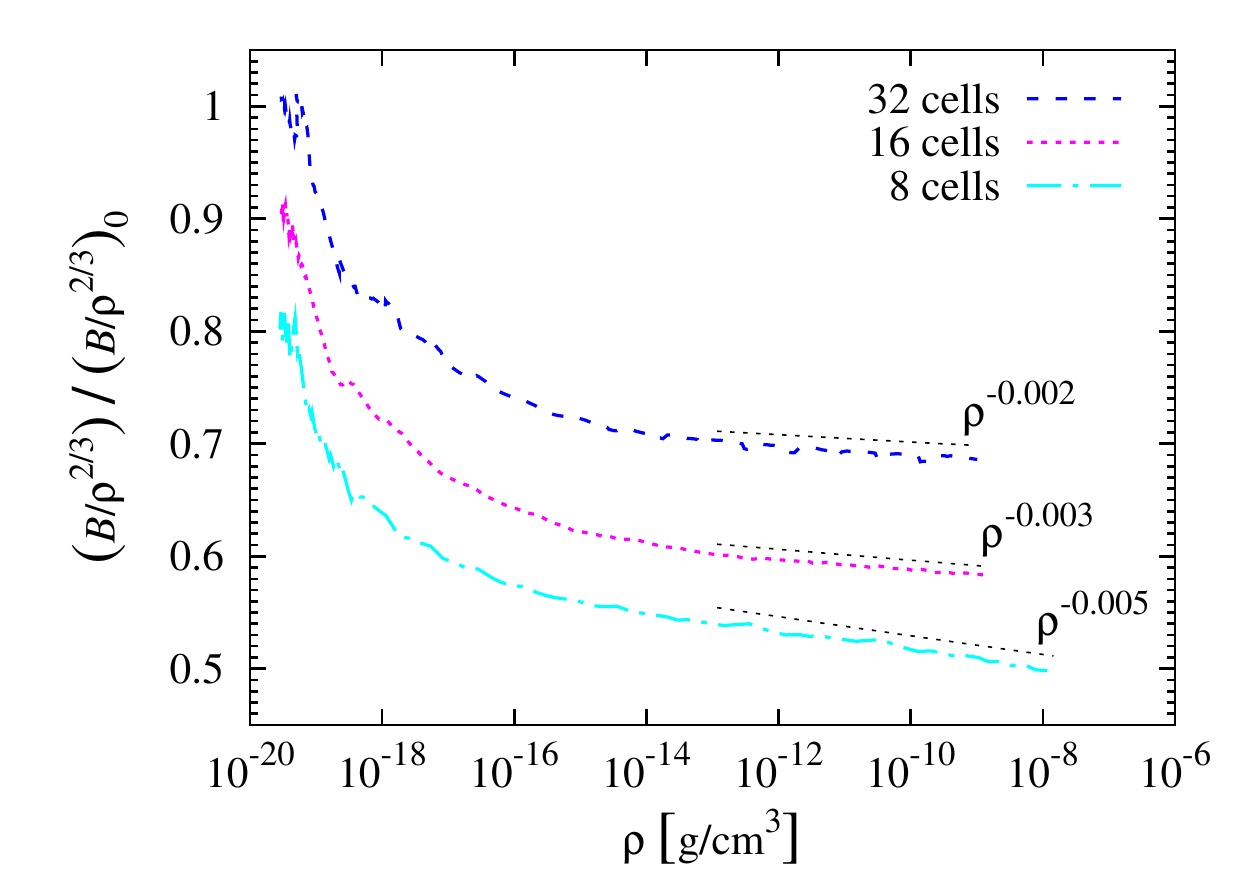}
\end{tabular}
\end{center}
\caption{Test on the numerical diffusion of the magnetic field in the HLL3R scheme for ideal MHD \citep{WaaganFederrathKlingenberg2011}. Three runs with Jeans resolutions of 8, 16, and 32 cells are shown. For clarity, the curves for 16 and 8 cells were shifted down by 10 and 20\%, respectively. The initial decrease of $B/\rho^{2/3}$ is due to the anisotropic geometry of the initial field structure. At late times during the collapse, the field becomes approximately isotropic due to the bending of the field lines to an hour-glass shape. In this regime, we can estimate upper limits for the effects of numerical diffusion, indicated by the dashed lines for $\rho>10^{-13}\,\g\,\cm^{-3}$.}
\label{fig:diffusion}
\end{figure}

In Figure~\ref{fig:diffusion} we show the magnetic field strength divided by $\rho^{2/3}$ as a function of the density, measured within the collapsing, central Jeans volume (see analysis described in section~\ref{sec:methods_analysis}). The magnetic field grows with increasing density, but initially, it grows shallower than $B\propto\rho^{2/3}$ due to the geometry of the field. Only for a fully isotropic field, $B\propto\rho^{2/3}$ for ideal MHD. Here, however, the magnetic field is initially very anisotropic (containing a $z$-component only), and thus, the magnetic field is initially only compressed along the $x-$, and $y$-directions, while the density increases along all three axes. Due to the bending of the field lines, the field becomes more isotropic in the late stages of the collapse. Figure~\ref{fig:diffusion} indeed shows a clear flattening of $B/\rho^{2/3}$ at high densities, when due to the bending, the magnetic field has developed roughly similar magnetic field strengths in all three directions. Thus, only then, we expect $B\propto\rho^{2/3}$ for ideal MHD. Power-law fits in this regime are indicated in Figure~\ref{fig:diffusion}. The curves are still falling with a roughly constant dependence on density, giving an upper limit on the effects of numerical diffusion. Three runs with Jeans resolutions of 8, 16, 32 cells were analyzed in this diffusion test. We find a departure from perfect flux-freezing, $B\propto\rho^{2/3-\epsilon}$ with $\epsilon\approx0.005$, $0.003$, and $0.002$ for a Jeans resolution of 8, 16, and 32 cells, respectively. Thus, the effects of numerical diffusion are resolution-dependent, as expected, however, the level of numerical diffusion of the magnetic field with respect to ideal MHD is generally quite small.

In this simple test, we only studied a particular situation, motivated by the density and magnetic field regimes analyzed in the main part of the paper. It should be noted, however, that the properties of numerical diffusion depend on the physical and numerical parameters of the modeled system.

\end{appendix}

\end{document}